\documentclass[reprint,superscriptaddress,nofootinbib,amsmath,amssymb,prd,floatfix]{revtex4-1}
\pdfoutput=1
\usepackage{graphicx}
\usepackage[dvipsnames]{xcolor}
\usepackage{amsmath}
\allowdisplaybreaks
\usepackage{amssymb}

\usepackage{slashed}
\usepackage[utf8]{inputenc}
\usepackage[colorlinks=true,linkcolor=blue,bookmarksopen,bookmarksnumbered]{hyperref}
\usepackage{color}
\usepackage[normalem]{ulem}
\usepackage{soul}
\usepackage{units}
\usepackage{rotating}
\usepackage{hhline,multirow,tabularx}
\usepackage{bm}
\usepackage{hyperref}
\usepackage[export]{adjustbox}
\usepackage{mdframed}

\newcommand{\diff}[1]{\mathrm{d}#1}

\def\bea#1\eea{\begin{align}#1\end{align}} 
\newcommand{\nnu}{\nonumber\\}
\newcommand{\bef}{\begin{figure}[htb]\centering}
\newcommand{\eef}{\end{figure}}

\newcommand{\nn}{\nonumber}

\def\L{\Lambda}

\def\<{\langle}
\def\>{\rangle}

\def\({\left(}
\def\[{\left[}
\def\){\right)}
\def\]{\right]}

\def\cos{\hbox{cos}}
\def\sin{\hbox{sin}}
\def\ln{\hbox{ln}}

\usepackage{lipsum}

\begin{document}

\title{Transverse Lambda production at the future Electron-Ion Collider}
	
\author{Zhong-Bo Kang}
\email{zkang@g.ucla.edu}
\affiliation{Department of Physics and Astronomy, University of California, Los Angeles, California 90095, USA}
\affiliation{Mani L. Bhaumik Institute for Theoretical Physics, University of California, Los Angeles, California 90095, USA}
\affiliation{Center for Frontiers in Nuclear Science, Stony Brook University, Stony Brook, New York 11794, USA}

\author{John Terry}
\email{johndterry@physics.ucla.edu}
\affiliation{Department of Physics and Astronomy, University of California, Los Angeles, California 90095, USA}
\affiliation{Mani L. Bhaumik Institute for Theoretical Physics, University of California, Los Angeles, California 90095, USA}

\author{Anselm Vossen}
\email{anselm.vossen@duke.edu}
\affiliation{Department of Physics, Duke University, Durham, North Carolina 27708, USA}
\affiliation{Jefferson Lab, 12000 Jefferson Avenue, Newport News, VA 23606, USA}

\author{Qinghua Xu}
\email{xuqh@sdu.edu.cn}
\affiliation{Institute of Frontier and Interdisciplinary Science \&  Key Laboratory of Particle Physics and Particle Irradiation (MoE),  
Shandong University, Qingdao, Shandong 266237, China}

\author{Jinlong Zhang}
\email{jlzhang@email.sdu.edu.cn}
\affiliation{Institute of Frontier and Interdisciplinary Science \&  Key Laboratory of Particle Physics and Particle Irradiation (MoE), Shandong University, Qingdao, Shandong 266237, China}
\affiliation{Center for Frontiers in Nuclear Science, Stony Brook University, Stony Brook, New York 11794, USA}

\begin{abstract}
We provide a comprehensive overview of transversely polarized $\L$ production at the future Electron-Ion Collider (EIC). In particular, we study both spontaneous transverse $\L$ polarization as well as the transverse spin transfer within the Transverse Momentum Dependent (TMD) factorization region. To describe spontaneous $\L$ polarization, we consider the contribution from the TMD Polarizing Fragmentation Function (TMD PFF). Similarly, we study the contribution of the transverse spin transfer originating from the transversity TMD Fragmentation Function (TMD FF). We provide projections for the statistical uncertainties in the corresponding spin observables at the future EIC. Using these statistical uncertainties, we characterize the role that the future EIC will play in constraining these distributions. We perform an impact study in the semi-inclusive deep inelastic scattering process for spontaneous $\L$ polarization with a proton beam. We find that the projected experimental data leads to a significant decrease in the uncertainties for the $u$ and sea TMD PFFs. Furthermore, to access the impact of the EIC on the transversity TMD FF, we perform the first extraction of the transversity TMD FF from the recent COMPASS data. We compare the statistical uncertainties of the future EIC with the theoretical uncertainties from our extraction and find that the EIC could have a significant role in constraining this distribution. Finally, we also provide projections for both spontaneous $\L$ polarization as well as the transverse spin transfer inside the jets in back-to-back electron-jet production at the EIC.
\end{abstract}
\maketitle
\section{Introduction}\label{Introduction}
Hadronization remains one of the most active and important areas of research in the field of nuclear physics. Within the past several decades, researchers have made tremendous advances in the understanding of Transverse Momentum Distribution Functions (TMDs), which allow us to uncover three-dimensional information for hadronization, as well as the three-dimensional structure of hadrons~\cite{Collins:1981uk, Collins:1981uw, Collins:1984kg, Ji:2004wu, GarciaEchevarria:2011rb, Collins:2011zzd}. In addition, these distributions allow us to understand correlations between the transverse momentum and the spin degrees of freedom. For recent reviews, see Refs~\cite{Metz:2016swz,Anselmino:2020vlp}. Over the past decade, there have been intense experimental and theoretical interests in understanding TMDs. In particular, a large number of phenomenological extractions have been performed for the unpolarized TMD Parton Distribution Functions (TMD PDFs), see for instance Refs.~\cite{Anselmino:2013lza,Bacchetta:2017gcc,Scimemi:2017etj,Bertone:2019nxa,Scimemi:2019cmh,Cammarota:2020qcw}. In addition, there has also been tremendous success in extracting spin-dependent TMDs, such as the Sivers function in Refs.~\cite{Echevarria:2014xaa,Echevarria:2020hpy,Cammarota:2020qcw,Bacchetta:2020gko,Bury:2020vhj,Bury:2021sue} and the transversity TMD PDFs in \cite{Kang:2015msa,Cammarota:2020qcw}. Despite the progress in understanding the spin-dependent TMD PDFs, probing the TMD Fragmentation Functions (TMD FFs) for polarized hadron production introduces additional complications due to experimental uncertainties associated with reconstructing the measured hadron's spin. As $\L$ and $\bar{\L}$ baryons, which we collectively denote $\L$ baryons in this paper, undergo self-analyzing weak decay, experimental measurements of $\L$ baryons plays a vital role in exploring spin-dependent TMD FFs.

One of the primary goals of the future Electron-Ion Collider (EIC) \cite{Accardi:2012qut,AbdulKhalek:2021gbh,Anderle:2021wcy} is to measure TMD FFs over wide kinematic regions at unprecedented experimental precision. In this paper, we aim to study the role that the future EIC can play in constraining TMD FFs which are associated with transversely polarized $\L$ production. However, the EIC could play a large role in constraining both longitudinal and transverse TMD FFs, see for instance Ref.~\cite{Chen:2021zrr}. One of two distributions that we study in this paper is the TMD Polarizing Fragmentation Function (TMD PFF), which characterizes the probability for an unpolarized quark to fragment into a transversely polarized $\L$. The Belle collaboration recently performed a measurement of spontaneous transverse $\L$ polarization in $e^+e^-$-annihilation in back-to-back $\L$ and light hadron $h=\pi^\pm,~ K^\pm$ production in Ref.~\cite{Guan:2018ckx}. This experimental data allowed for the first phenomenological extractions of the TMD PFF in \cite{DAlesio:2020wjq,Callos:2020qtu} within the TMD factorization formalism~\cite{Collins:2011zzd}, as well as a follow up extraction in~\cite{Chen:2021hdn}. The second TMD FF which we study in this paper is the transversity TMD FF, which characterizes the probability for a transversely polarized quark to fragment into a transversely polarized $\L$ baryon. The COMPASS collaboration performed recent measurements of the transverse spin transfer in Semi-Inclusive Deep Inelastic Scattering (SIDIS) for $\L$ production in Ref.~\cite{Alexeev:2021aws}, which can be naturally studied via the TMD factorization formalism. This experimental measurement opens the possibility of performing the first extraction of the quark transversity TMD FF for $\L$ production. In addition, the STAR experiment also reported their measurements on transverse spin transfer for single inclusive $\Lambda$/$\bar\Lambda$ hyperon production in proton-proton collisions at $\sqrt s=$ 200 GeV \cite{STAR:2018fqv}. However, while the STAR measurement can be used as a probe of the collinear transversity PDF and transversity FF, the process is described by the collinear factorization formalism~\cite{deFlorian:1998am,Collins:1989gx} and not by the TMD factorization formalism.

While the future EIC offers the possibility of measuring spontaneous $\L$ polarization and the transverse spin transfer in SIDIS, recently back-to-back electron-jet production in electron-proton, $e+p$, collisions has been explored as a probe of the TMD PDFs in Refs.~\cite{Liu:2018trl, Liu:2020dct}. Furthermore, in Refs.~\cite{Kang:2017glf, Kang:2019ahe, Kang:2021ffh} the authors discuss that by measuring the distribution of hadrons relative to the jet axis, one de-correlates the TMD FF in the TMD fragmenting jet function and the other TMDs in the process. As a result, in Ref.~\cite{Arratia:2020nxw, Kang:2021ffh}, it was proposed to measure the distribution of hadrons in a jet in back-to-back electron-jet production as a probe of TMD FFs. 

To address the role that the future EIC can play in constraining the TMD PFF, in this paper we perform an EIC impact study for the SIDIS process in extracting the TMD PFF. Furthermore, we use the recent COMPASS measurement for the transverse spin transfer to perform the first extraction of the quark-to-$\L$ transversity TMD FF. Using this extraction, we compare our theoretical uncertainties against our projections for the statistical uncertainties at the future EIC. Finally, we also provide projections at the future EIC for back-to-back electron-jet production for both spontaneous $\L$ polarization in unpolarized $ep$ collisions, as well as the transverse spin transfer in transversely polarized $ep$ scattering.

The paper is organized as follows. In Sec.~\ref{sec:Theory}, we provide the theoretical formalism for spin configuration in the SIDIS process as well in back-to-back electron-jet production. In Sec.~\ref{Sec:Exp}, we provide the details for the simulated experimental setup. In Sec.~\ref{sec:Simulation}, we provide the details and results of our EIC impact study for the TMD PFF. In Sec.~\ref{sec:SIDIS-Spin-Transfer}, we provide the details for our extraction of the transversity TMD FF. In Sec.~\ref{sec:In-Jet} we provide our projections for spontaneous $\L$ in jet polarization as well as the transverse spin transfer. We summarize our findings and conclude in Sec.~\ref{sec:Conclusion}.

\section{QCD factorization}\label{sec:Theory}
In this section, we first review the TMD factorization formalism for spontaneous $\L$ polarization as well as the transverse spin transfer in SIDIS. We then provide the factorization formalism for $\L$ production inside the jet in back-to-back electron-jet production in $ep$ collisions, with which we study two aforementioned spin configurations. We demonstrate that the spontaneous $\L$ polarization allows us to probe TMD PFF, while the transverse spin transfer is sensitive to $\L$ transversity TMD FF.

\subsection{SIDIS}\label{subsec:SIDIS-theory}
We start with the production of $\L$ baryons in SIDIS,
\begin{align}
    e(\ell)+p(P,\bm{s}_{\perp})\rightarrow e(\ell^\prime)+\L(P_h,\bm{s}_{\Lambda \perp})+X \,,\nn
\end{align} 
where $\bm{s}_{\perp}$ is the transverse spin of the incoming proton, while $\bm{s}_{\Lambda \perp}$ is the transverse spin of the final produced $\L$ baryon. We take the frame choice such that the proton moves in the positive $z$ direction while the incoming virtual photon moves in the negative $z$ direction, see Fig.~1 of Ref.~\cite{Kang:2015msa} for our convention, alternatively see Ref.~\cite{Bacchetta:2004jz} for the so-called Trento conventions. In the proton-photon COM frame, the differential cross section can be written as
\begin{align}
\frac{d\sigma(\bm{s}_{\perp},\bm{s}_{\Lambda \perp})}{d\mathcal{PS}\, d^2 P_{h \perp}} & = \sigma_0^{\rm DIS} \bigg[ F_{UU} \\
& + \sin(\phi_{S} - \phi_\L) F_{UT}^{\sin(\phi_{S} - \phi_\L)} \nn \, \\
& + \cos{\left(\varphi_{S}-\phi_{S} \right)}\,D(y)\,
F_{TT}^{\cos\left(\varphi_{S}-\phi_{S}\right)} \bigg],\nn
\label{e.sidis}
\end{align}
where $d\mathcal{PS} = dx_B\, dy\, dz_\L$ is the phase space element associated with the SIDIS differential cross section while $d^2 P_{h \perp}$ is the phase space element associated with the transverse momentum of the $\L$ baryon. Additionally
\begin{align}
x_B = \frac{Q^2}{2P\cdot q},\qquad
y = \frac{P\cdot q}{P\cdot \ell},\qquad
z_\L = \frac{P\cdot P_h}{P\cdot q},
\end{align}
are the standard SIDIS kinematic variables and $Q^2 = -q^2 = - (\ell' - \ell)^2$. At the same time
\begin{align}
\sigma_0^{\mathrm{DIS}} &= \frac{2\pi\alpha_{\mathrm{em}}^2}{Q^2}\frac{1+(1-y)^2}{y}\,,
\\
D(y) &= \frac{2(1-y)}{1+(1-y)^2}\,.
\end{align}
In Eq.~\eqref{e.sidis}, the $F$ terms denote the structure functions, where the first subscript denotes the polarization of the incoming proton while the second subscript denotes the polarization of the outgoing $\L$. On the other hand, the terms $\varphi_S$ and $\phi_S$ in the superscript of the $F_{TT}^{\cos(\varphi_S - \phi_S)}$ structure function denote the azimuthal angles for $\bm{s}_{\perp}$  and $\bm{s}_{\Lambda \perp}$, respectively. Furthermore  the $\phi_\L$ term in the superscript of the $F_{UT}^{\sin(\phi_S-\phi_\L)}$ structure function denotes the azimuthal angle of the  transverse momentum of the $\L$ baryon, which is denoted $\bm{P}_{h \perp}$. 

The experimentally measured spontaneous transverse polarization $P_{\L}$ and the transverse spin transfer $S_\L$ for $\L$ production are given by
\begin{align}\label{eq:pol-def}
    P_{\L} =&\, \frac{F_{UT}^{\sin(\phi_S-\phi_\L)}}{F_{UU}}\,,
    \nnu
    S_\L =&\, D(y)\,\frac{F_{TT}^{\cos\left(\varphi_{S}-\phi_{S}\right)}}{F_{UU}}\,, 
\end{align}
respectively. We note at this point that the COMPASS measurement in Ref.~\cite{Alexeev:2021aws} did not include the depolarization factor $D(y)$ in the definition of the transverse spin transfer. 

Within the usual TMD factorization, the structure functions can be written as
\begin{align}
\label{eq:strUU}
F_{UU} & =  H^{\rm DIS}(Q)\,\mathcal{F} \left[f\, D \right]\,,
\\
\label{eq:strUT}
F_{UT}^{\sin(\phi_S - \phi_\L)} & =
H^{\rm DIS}(Q)\,\mathcal{F}\left[\frac{\hat{\bf{P}}_{h \perp}\cdot \bm{p}_{\perp}}{z_\L M_\L} f\,D_{1T}^{\perp}\right]\,,
\\
\label{eq:strTT}
F_{TT}^{\cos(\varphi_S - \phi_S)} & =
H^{\rm DIS}_\perp(Q)\,\mathcal{F}\left[ h\, H \right]\,.
\end{align}
In these expressions $\mathcal{F}$ denotes the convolution integral which is defined as
\begin{align}
\label{e.Fcov-dis}
\mathcal{F}\left[c\,A\,B \right] &= \sum_q e_q^2 \int d^2 \bm{k}_{\perp} d^2 \bm{p}_{\perp} \\
& \hspace{-6mm}\times \delta^{(2)}(z_\L \bm{k}_{\perp}+\bm{p}_{\perp} -\bm{P}_{h \perp}) \nonumber \\
&\hspace{-6mm} \times c(\bm{k}_\perp, \bm{p}_{\perp})\,A_{q/p}(x_B, k_{\perp}^2, Q)\, B_{\L/q}(z_\L, p_{\perp}^2, Q)\, \nn.
\end{align}
Here $\bm {k}_{\perp}$ denotes the transverse momentum of the quark relative to the initial proton, $\bm {p}_{\perp}$ denotes the transverse momentum of the $\L$ relative to the parent quark, and $\hat{\bm{P}}_{h \perp} = \bm{P}_{h \perp}/|\bm{P}_{h \perp}|$. Additionally $c(\bm{k}_\perp, \bm{p}_{\perp})$ represents arbitrary functions of $\bm {k}_{\perp}$, $\bm {p}_{\perp}$, and $\hat{\bm{P}}_{h \perp}$ that enter into the cross sections in Eqs.~\eqref{eq:strUU}, \eqref{eq:strUT}, and \eqref{eq:strTT}, respectively. 
For example, for Eq.~\eqref{eq:strUT}, we identify 
\bea
c(\bm{k}_\perp, \bm{p}_{\perp}) = \frac{\hat{\bf{P}}_{h \perp}\cdot \bm{p}_{\perp}}{z_\L M_\L}\,.
\eea
The functions $H^{\rm DIS}(Q)$ and $H^{\rm DIS}_\perp(Q)$ in these expressions denote the hard functions for the unpolarized and transversely polarized quark channels. We normalize both of these functions to 1 at Leading Order (LO). To arrive at Eqs.~\eqref{eq:strUU}, \eqref{eq:strUT}, and \eqref{eq:strTT}, we have made the scale choice $\mu^2 = \zeta = Q^2$ where $\mu$ is the renormalization scale while $\zeta$ is the rapidity scale for the TMD PDF and TMD FF in the factorization formalism~\cite{Collins:2011zzd}. Furthermore, in Eqs.~\eqref{eq:strUU}, \eqref{eq:strUT}, and \eqref{eq:strTT} we introduce the relevant TMDs for each of the possible spin configurations. In these expressions, $f_{q/p}$ and $D_{\L/q}$ are the unpolarized TMD PDF and unpolarized TMD FF, $D_{1T\,\L/q}^{\perp}$ is the TMD PFF, and $h_{q/p}$ and $H_{\L/q}$ are the transversity TMD PDF and TMD FF, respectively. It is worthwhile noting that there is another term which contributes to the $\L$ transverse polarization in unpolarized $ep$ collisions and results in a $\sin(\phi_S+\phi_\L)$-azimuthal modulation. This contribution arises from the Boer-Mulders function in the proton convoluted with the transversity TMD FF~\cite{Zhou:2008fb}. 

At this point, we note that the four dimensional convolution integrals in Eq.~\eqref{e.Fcov-dis} can be  simplified to a single integral by making the replacement
\begin{align}
    \delta^{(2)}& (z_\L \bm{k}_{\perp}+\bm{p}_{\perp} -\bm{P}_{h \perp}) \\
    & = \frac{1}{z_\L^2} \int \frac{d^2 b}{(2\pi)^2} e^{-i \bm{b}\cdot(\bm{k}_\perp+\bm{p}_\perp/z_\L-\bm{P}_{h\perp}/z_\L)}\nn \,.
\end{align}
After making this replacement, the azimuthal angle of $\bm{b}$ can be integrated over. The unpolarized structure function for example then has the simple form
\begin{align}\label{eq.strFUU}
    F_{UU} = H^{\rm DIS}(Q)\,\sum_q e_q^2 \int & \frac{db\, b}{2\pi} \,J_0\left(\frac{bP_{h \perp}}{z_\L}\right) \\
    \times & D_{\L/q}(z_\L,b, Q) f_{q/p}(x_B,b, Q)\,  \nn
\end{align}
where $b = |\bm{b}|$ and
\begin{align}
    f(x_B,b,Q) = \int d^2\bm{k}_T e^{-i \bm{b}\cdot k_\perp} f(x_B,k_\perp,Q),
\end{align}
\begin{align}
    D(z_{\Lambda},b,Q) = \frac{1}{z_{\Lambda}^2}\int d^2\bm{p}_\perp e^{-i \bm{b}\cdot p_\perp/z_{\Lambda}} D(z_{\Lambda},p_\perp,Q),
\end{align}
represent the Fourier transforms of the unpolarized TMD PDF and TMD FF. Looking at Eq.~\eqref{eq.strFUU}, one then regards $\bm{b}$ as the Fourier conjugate vector to $\bm{q}_\perp = -\bm{P}_{h\perp}/z_\L$. The polarized structure functions in Eqs.~\eqref{eq:strUT} and \eqref{eq:strTT} can also be written in $b$-space as a single integral over $b$. 

\subsection{$\L$ inside a jet}\label{subsec:In-Jet-theory}
We will now discuss the factorization formalism for transverse $\L$ production inside a jet for the back-to-back electron-jet production in $ep$ collisions
\begin{figure*}[hbt!]
    \centering
    \includegraphics[width = 0.8\textwidth]{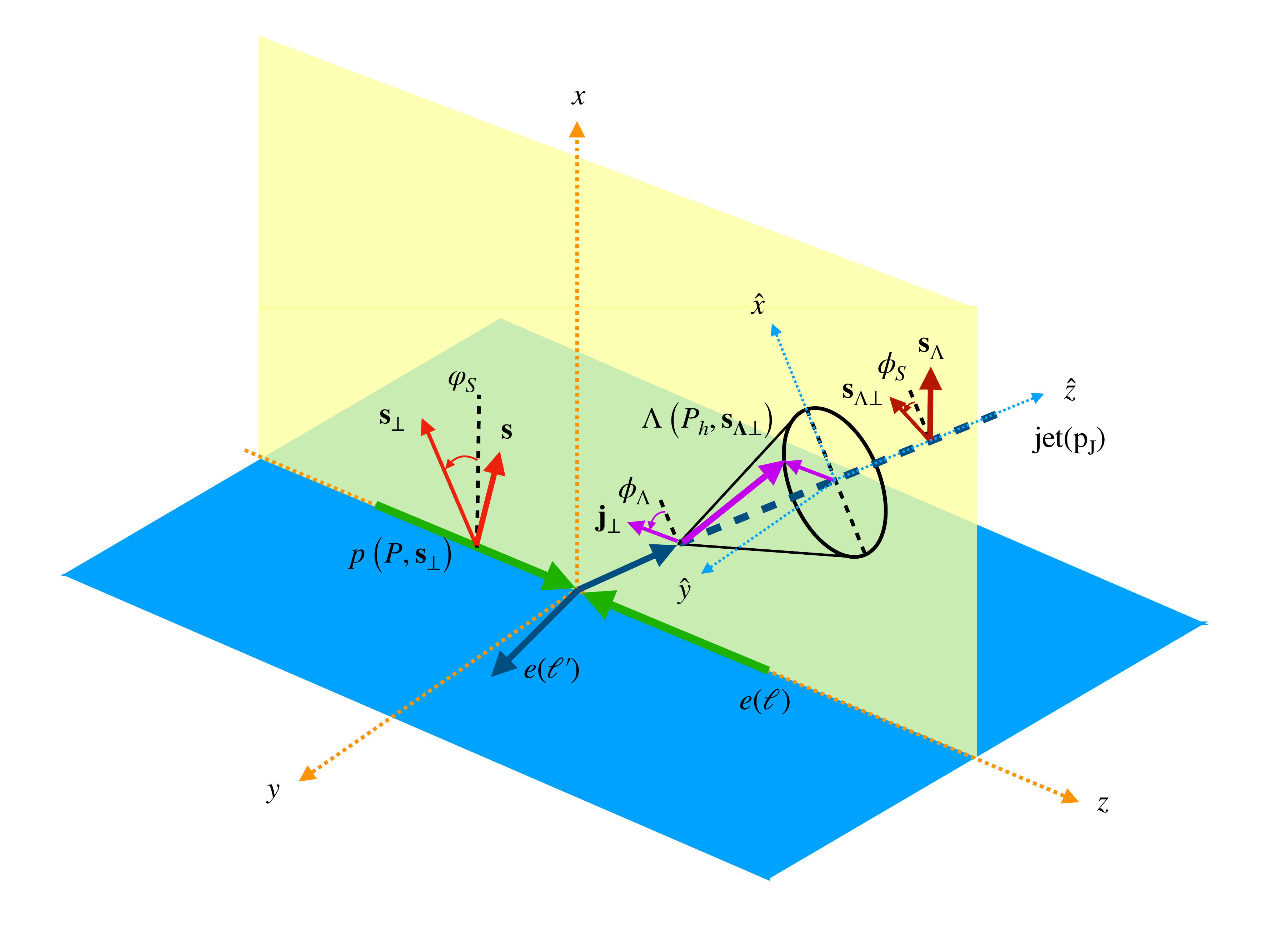}
    \caption{Kinematic configuration for back-to-back lepton-jet production.}
    \label{fig:jet-l}
\end{figure*}
\begin{align}
e(\ell)+p(P,\bm{s}_\perp)\to e(\ell')+({\rm jet}(p_{J})\,\L(P_h, \bm s_{\L \perp}))+X \nn \,. 
\end{align}
In Fig.~\ref{fig:jet-l}, we have included a plot which demonstrates the kinematic configuration of this process. The jet is constructed via a proper jet algorithm such as anti-$k_T$ algorithm~\cite{Cacciari:2011ma} with the jet radius $R$. For this process, we denote the transverse momentum of the jet direction as $\bm{p}_{J\perp}$ while $\bm{\ell}'_\perp$ represents the transverse momentum of the final state lepton. In both cases, the transverse momenta are defined in the center-of-mass frame of the incoming electron and the incoming nucleon. In this frame $\bm{q}_\perp=\bm \ell'_\perp + \bm p_{J\perp}$ represents the transverse momentum imbalance of the outgoing electron and the jet. The back-to-back electron-jet configuration occurs at a small transverse momentum imbalance, $|\bm{q}_\perp| \ll \ell'_\perp \sim p_{J\perp}$~\cite{Liu:2018trl,Arratia:2020nxw}. Additionally, for this process it is convenient to measure the transverse momentum of the $\L$ baryon relative to the jet axis, which we denote $\bm{j}_\perp$. The TMD region for $\L$ production in the jet occurs in the kinematic region where $j_\perp \ll p_{J\perp} R$~\cite{Kang:2017glf}. 

By studying the partonic process for jet production, one can see that the direction of the jet is directly sensitive to the TMD PDF. By measuring the transverse momenta of $\L$ baryons within this jet, we also gain sensitivity to the TMD FFs. This process offers the advantage that since the jet and the $\L$ baryons are measured with respect to different axes, the factorization structure for this process will lead to a deconvolution of the TMD PDF and TMD FF, thus allowing us to probe these structures more independently.

Following the work of~\cite{Kaufmann:2015hma,Kang:2016ehg,Bain:2016rrv,Kang:2017glf,Kang:2017btw,Kang:2019ahe,Kang:2020xyq}, the relevant cross section can be written as
\begin{align}\label{eq:structure}
    & \frac{\diff \sigma(\bm{s}_\perp,\bm{s}_{\L \perp})}{d{\cal PS}\,d^2 \bm q_\perp\,d z_{J \L}\,d^2 \bm j_\perp}  = 
    \sigma_0 \bigg[W_{UU} \\
    & \hspace{1cm} + \sin(\phi_S-\phi_\L) \,W_{UT}^{\sin(\phi_S - \phi_\L)} \nn\\
    & \hspace{1cm} + \cos\left(\varphi_{\mathcal{S}}-\phi_S\right)\,D(\hat{s},\hat{t},\hat{u})\,W_{TT}^{\cos\left(\varphi_{\mathcal{S}}-\phi_S\right)} \bigg]\,, \nn
\end{align}
In this expression $d{\cal PS} = dy_e\, d^2\bm \ell'_\perp$ is the phase space element in rapidity and transverse momentum of the outgoing electron in the center-of-mass frame of incoming electron and the nucleon. The variable $z_{J \L}$ represents the fraction of the momentum of the jet which is carried by the $\L$ baryon. The $z_{J \L}$ and $\bm{j}_\perp$ variables can be related to the momenta of the $\L$ baryon and jet through the relations
\begin{align}
z_{J \L} =& \bm{P}_h \cdot \bm{p}_{J}/|\bm{p}_{J}|^{2}\,,
\qquad
\bm j_\perp =& \bm{P}_h \times \bm{p}_{J}/|\bm{p}_{J}|\,,
\end{align}
where $\bm{P}_h$ and $\bm{p}_J$ represent the three momenta of the $\L$ baryon and the jet respectively.

On the other hand, in Eq.~\eqref{eq:structure}, we have
\begin{align}
    \sigma_0 = \frac{\alpha_{\rm em}^2}{s Q^2} \frac{2(\hat{s}^2+\hat{u}^2)}{Q^4}\,,
\end{align}
where the partonic Mandelstam variables are given by
\begin{align}
    & \hat{s} = x_B S_{\rm ep}\,,
    \qquad
    \hat{t} = -\sqrt{S_{\rm ep}}\,\ell_\perp' e^{y_e}\,,\\
    & \hat{u} = -x_B\sqrt{S_{\rm ep}}\,\ell_\perp' e^{-y_e}\,,\nn 
\end{align}
where $S_{\rm ep}$ is the center of mass energy of the electron-proton pair while $Q^2 = -\hat{t}$. Furthermore, we define
\begin{align}
    D(\hat{s},\hat{t},\hat{u}) = -\frac{2\hat{s}\hat{u}}{\hat{s}^2+\hat{u}^2}\,.
\end{align}
For this process, $W$ denote the structure functions and we follow the same labelling convention for the subscripts as outlined in the previous section. The polarization and spin transfer for back-to-back electron-jet production in $ep$ collisions can be written in terms of these structure functions as
\begin{align}
\label{eq:pol-in-jet-def}
    P_{\L} =&\, \frac{W_{UT}^{\sin(\phi_S-\phi_\L)}}{W_{UU}}\,,
    \\
\label{eq:trans-in-jet-def}
    S_\L =&\, D(\hat{s},\hat{t},\hat{u})\,
    \frac{W_{TT}^{\cos\left(\varphi_{S}-\phi_{S}\right)}}{W_{UU}}\,,
\end{align}
respectively.

The expression for the unpolarized structure function can be obtained from Ref.~\cite{Arratia:2020nxw} as
\begin{align}\label{eq:unp-jet-L}
W_{UU} &= \,H(Q,\mu) \sum_q e_q^2\, {\cal G}_{\L/q}(z_{J \L},j_\perp,\mu_J,\mu)
    \\ &\times \,
\int\frac{d^2 b}{(2\pi)^2}\, e^{i\bm q_\perp\cdot \bm b}\, f_{q/p}(x_B,b,\mu)\, U(b,y_{J},R,\mu) \,, \nonumber\,.
\end{align}
In this expression, $H$ and $U$ are the hard function and the $b$-space soft functions for this process. Furthermore, $x_B$ is the usual Bjorken variable which is related to the variables in the phase space element through the relation
\begin{align}
    x_B =  \frac{Q^2}{2 P\cdot q} = \frac{\ell_\perp' e^{y_e}}{\sqrt{S_{\rm ep}}-\ell_\perp' e^{-y_e}}\,.
\end{align}
For later convenience, we also define the inelasticity $y$ as
\begin{align}\label{eq:yey}
    y = 1-\frac{\ell_\perp'}{\sqrt{S_{\rm ep}}}e^{-y_e}\,.
\end{align}
The rapidity of the jet, $y_J$, can also be defined in terms of the kinematic variables entering into the phase space through the relation
\begin{align}
    y_{J} = -\frac{1}{2}\ln\left(\frac{\hat{t}}{x\hat{u}}\right).
\end{align}
Furthermore, $\mu$ and $\mu_J$ are the renormalization and the jet scales, respectively. For the remainder of this paper, we will always choose the renormalization scale to be given by $\mu=p_{J\perp}$ while the jet scale will be given by $\mu_J=p_{J\perp}R$. The function ${\cal G}_{\L/q}$ entering into Eq.~\eqref{eq:unp-jet-L} is the TMD fragmenting jet function ~\cite{Procura:2009vm,Kang:2017glf,Kang:2019ahe,Kang:2020xyq}, which describes the distribution of $\L$ particles inside the jet. Following the results of Ref.~\cite{Arratia:2020nxw}, this distribution function is related to the usual unpolarized TMD FF through the relation 
\begin{align}\label{eq:TMDFF}
    & {\cal G}_{\L/q}(z_{J \L},j_\perp,\mu_J,\mu)=\exp\left[\int_{\mu_J}^{\mu} \frac{d\mu'}{\mu'} \gamma_J(\mu') \right] \\
    & \hspace{1cm}\times \int\frac{d^2 b}{(2\pi)^2}e^{i\bm j_\perp\cdot\bm b/z_{J \L}} D_{\L/q}(z_{J \L}, b,\mu_J) \nn \,,
\end{align}
where $\gamma_J$ is the anomalous dimension of the TMD fragmenting jet function, to be given below in Sec.~\ref{sec:In-Jet}.

For spontaneous $\L$ polarization, we follow the procedure in Ref.~\cite{Gamberg:2021iat} to replace the TMD FF in Eq.~\eqref{eq:TMDFF} by the relevant density associated with the distribution of transversely polarized $\L$ baryons. Explicitly, we make the replacement
\begin{align}
    D_{\L/q}(z_{J \L}, b,\mu_J) \rightarrow -\frac{M_\L \epsilon_{\perp\rho\sigma} b^\rho S_{\perp}^{ \sigma}}{z_{J \L}^2} D_{1T,\L/q}^{\perp(1)}\left(z_{J \L},b,\mu_J\right)\,,\nn
\end{align}
where $D_{1T,\L/q}^{\perp(1)}$ is the first moment of the TMD PFF in $b$-space~\cite{Gamberg:2021iat}.  After making this replacement, the structure function for spontaneous $\L$ polarization is given by
\begin{align}\label{eq:pol-jet-L} 
&W_{UT}^{\sin(\phi_S - \phi_\L)} = H(Q,\mu) \sum_q e_q^2\, {\cal G}_{1T,\L/q}^{\perp}(z_{J \L},j_\perp,\mu_J,\mu) \nn \\
& \hspace{0.5cm} \times \int\frac{d^2 b}{(2\pi)^2}\, e^{i\bm q_\perp\cdot \bm b}\, f_{q/p}(x_B,b,\mu)\, U(b,y_{J},R,\mu)\,,
\end{align}
where
\begin{align}\label{eq:TMDPFF}
    & {\cal G}_{1T,\L/q}^{\perp}(z_{J \L},j_\perp,\mu_J,\mu)=\frac{M_\L}{z_{J \L}} \exp\left[\int_{\mu_J}^{\mu} \frac{d\mu'}{\mu'} \gamma_J(\mu') \right]  \nn
    \\
    & \hspace{0.5cm} \times \frac{\partial}{\partial j_\perp} \int\frac{d^2 b}{(2\pi)^2}e^{i\bm j_\perp\cdot\bm b/z_{J \L}} \, D_{1T,\L/q}^{\perp(1)}\left(z_{J \L},b,\mu_J\right) \,.
\end{align}

Finally, we define the transverse spin transfer structure function as
\begin{align}\label{eq:trans-jet-L}
&W_{TT}^{\cos\left(\varphi_{S}-\phi_{S}\right)} = \,H_\perp(Q,\mu) \sum_q e_q^2\, {\cal G}_{\L/q}^T(z_{J \L},j_\perp,\mu_J,\mu)\nn
    \\ &\times \,
\int\frac{d^2 b}{(2\pi)^2}\, e^{i\bm q_\perp\cdot \bm b}\, h_{q/p}(x_B,b,\mu)\, U(b,y_{J},R,\mu) \,,
\end{align}
where
\begin{align}\label{eq:TMDHFF}
    {\cal G}_{\L/q}^T&(z_{J \L},j_\perp,\mu_J,\mu)=\exp\left[\int_{\mu_J}^{\mu} \frac{d\mu'}{\mu'} \gamma_J(\mu') \right] \nonumber\\
    & \times \int\frac{d^2 b}{(2\pi)^2}e^{i\bm j_\perp\cdot\bm b/z_{J \L}} H_{\L/q}(z_{J \L}, b,\mu_J)\,, 
\end{align}
provides the distribution of transversely polarized $\L$ baryons in a jet which is initiated by a transversely polarized quark and $H_\perp(Q, \mu)$ is the hard function associated with the transversely polarized quark hard process.
\section{Experimental Setup}\label{Sec:Exp}
In this section, we present the details of our simulation for generating the event statistics. The present study is based on the four baseline energy configurations which are discussed in EIC Yellow Report~\cite{khalek:2021science}. The four configurations are 5 GeV $\times$ 41 GeV, 5 GeV $\times$ 100 GeV, 10 GeV $\times$ 100 GeV, and 18 GeV $\times$ 275 GeV, where the first energy is the electron beam energy while the second energy is the proton beam energy. The $ep$ event simulation that we present here is based on the PYTHIA eRHIC Monte Carlo program which is a modified version of PYTHIA-6.4.28~\cite{Sjostrand:2006za} with the PDFs input from the LHAPDF~\cite{Whalley:2005nh} library. Furthermore, for the back-to-back lepton-jet process, we perform jet reconstruction using the FASTJET~\cite{Cacciari:2011ma} package. The kinematics have been constrained in the following ranges: $Q>1$ GeV, $0.05<y<0.95$, $W>2$ GeV. The constrains on $Q^2$ and $W$ are used to select valid SIDIS events, whereas the $y$ selection avoids phase space where either radiative corrections become large or the event cannot be reliably reconstructed.

\begin{figure}[hbt!]
    \centering
    \includegraphics[width = 0.48\textwidth]{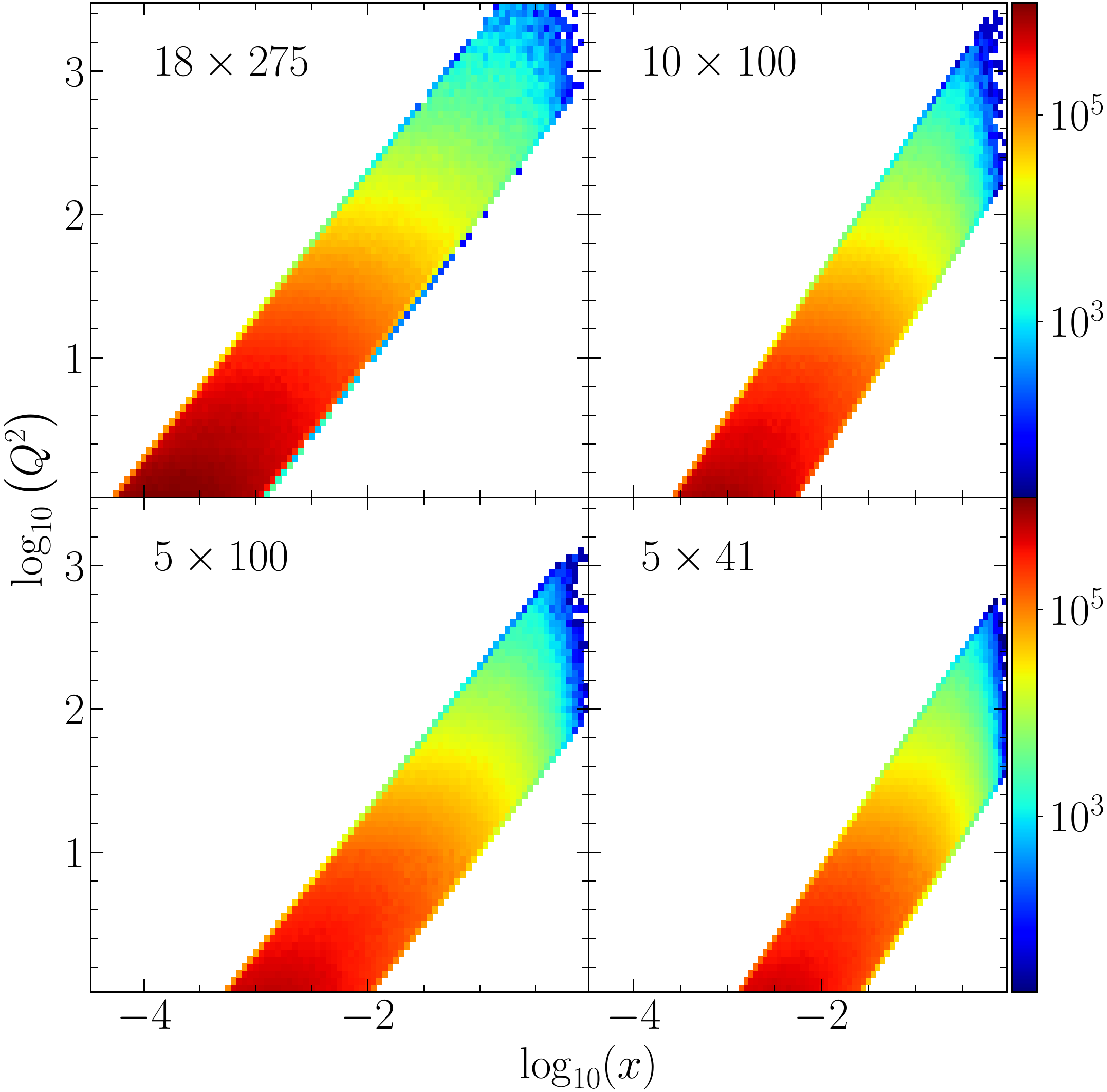}
    \caption{The ranges in the square of the transferred photon momentum $Q^2$ versus the parton momentum fraction $x$ accessible for different collision energies. The $z$-scale (density) indicates the number of events with at least one $\Lambda$/$\bar{\Lambda}$ in pseudo-rapidity range -3.5$<\eta<$3.5.}
    \label{fig:x_q2}
\end{figure}

\begin{figure*}[hbt!]
    \centering
    \hfill
    \includegraphics[height = 2in]{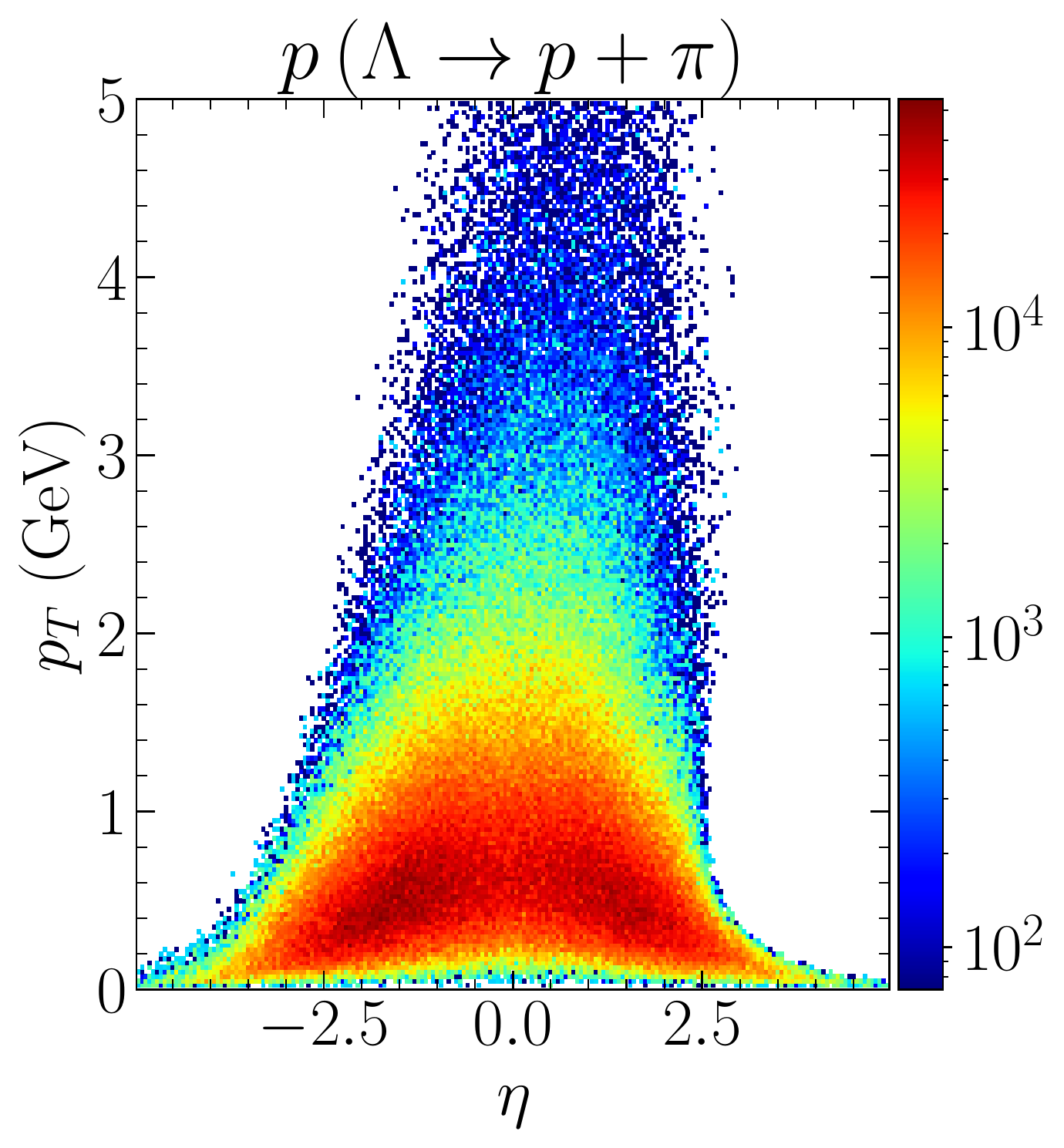}
    \hfill
    \includegraphics[height = 2in]{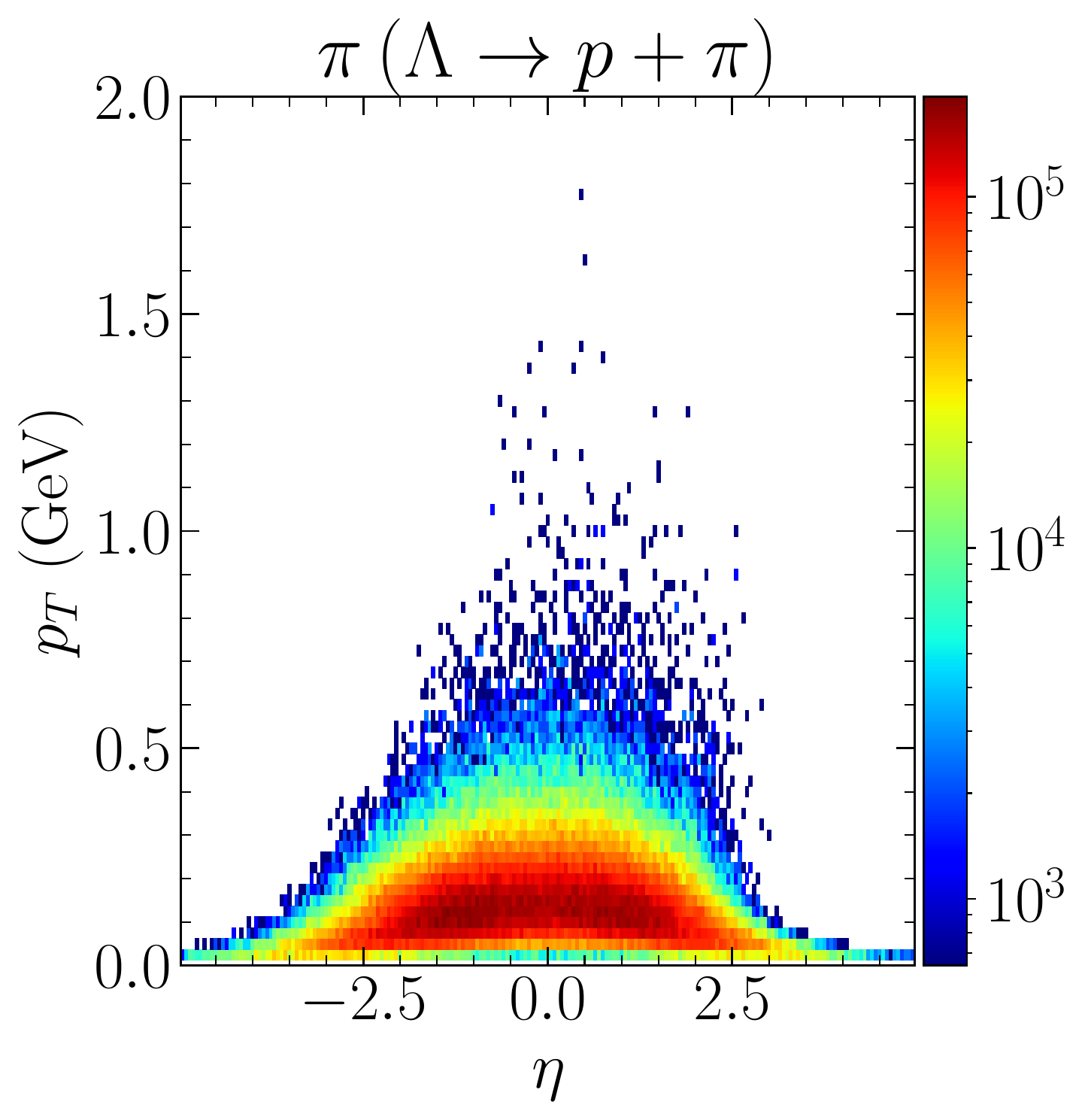}
    \hfill
    \includegraphics[height = 2in]{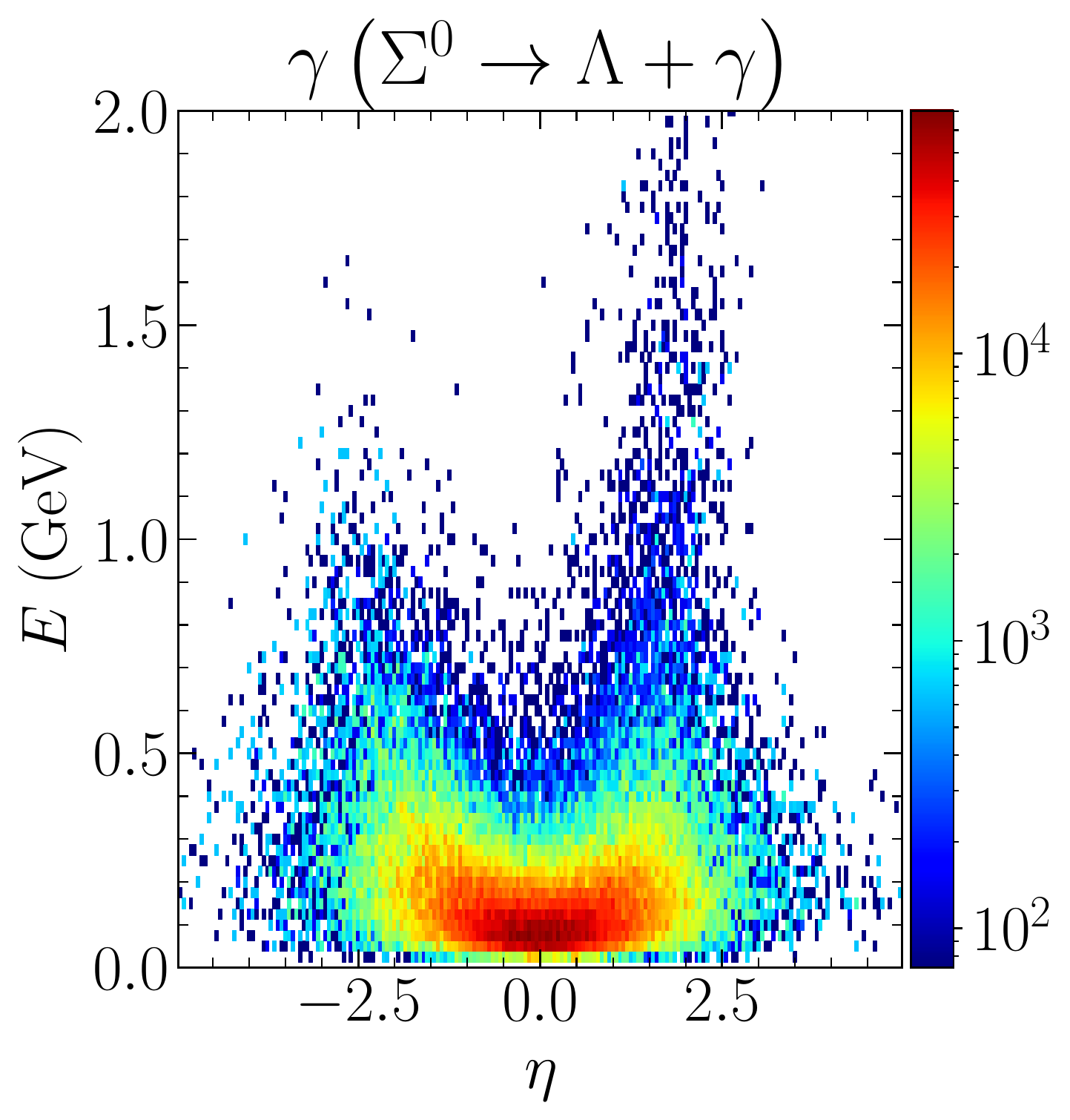}
    \hfill
    \caption{Final state particle distributions in transverse momentum (energy) and pseudo-rapidity space for proton (left), pion (middle), and photon from $\Sigma^0$ decay (right). Here we display the results at the collision energy 18$\times$275 GeV$^2$ and we note that the distributions for other energy configurations are similar.}
    \label{fig:daughter_pt_eta}
\end{figure*}

\begin{figure}[hbt!]
    \centering
    \includegraphics[width = 0.4\textwidth]{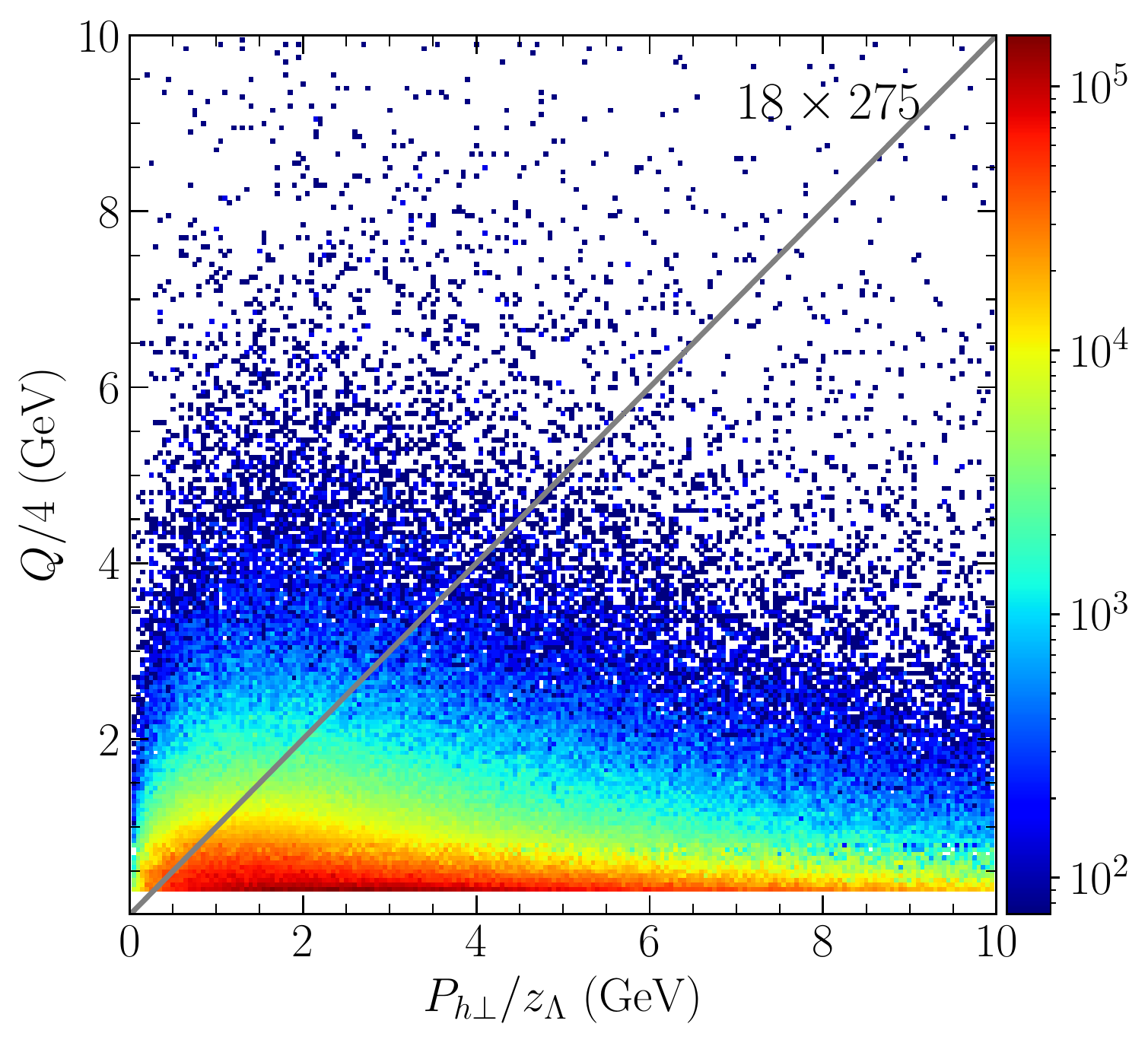}
    \caption{Illustration of the impact of the constraint $P_{h \perp}/z_\L < Q/4$. Only events above the line are accepted. The efficiency of this cut depends on the hard scale $Q^2$ and is approximately 7\% for $Q^2>1$~GeV$^2$, 25\% for $Q^2> 10$~GeV$^2$ and 50\% for $Q^2>100$~GeV$^2$. Here, we provide the result for the 18$\times$275 GeV$^2$ energy configuration while we note that the results for the other collision energies configurations are similar.}
    \label{fig:lambda_qtZ}
\end{figure}

\begin{figure}[hbt!]
    \centering
    \includegraphics[width = 0.35\textwidth]{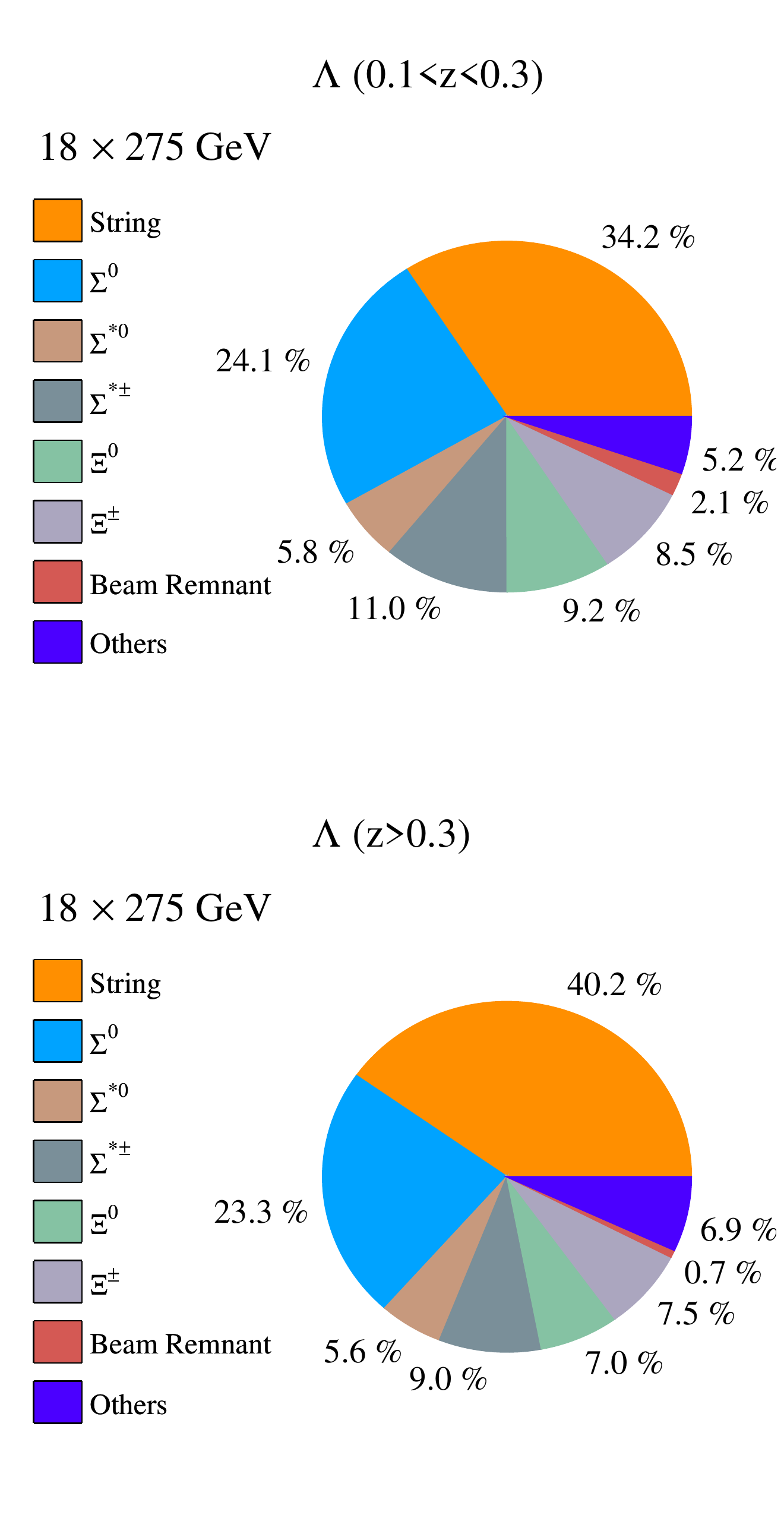}
    \caption{Origins of $\L$ for $0.1<z<0.3$ and $z>0.3$ according to the event records provided by the PYTHIA event generator for the 18$\times$275 GeV$^2$ energy configuration.}
    \label{fig:lambda_origins}
\end{figure}

\begin{figure}[hbt!]
    \centering
    \includegraphics[width = 0.49\textwidth]{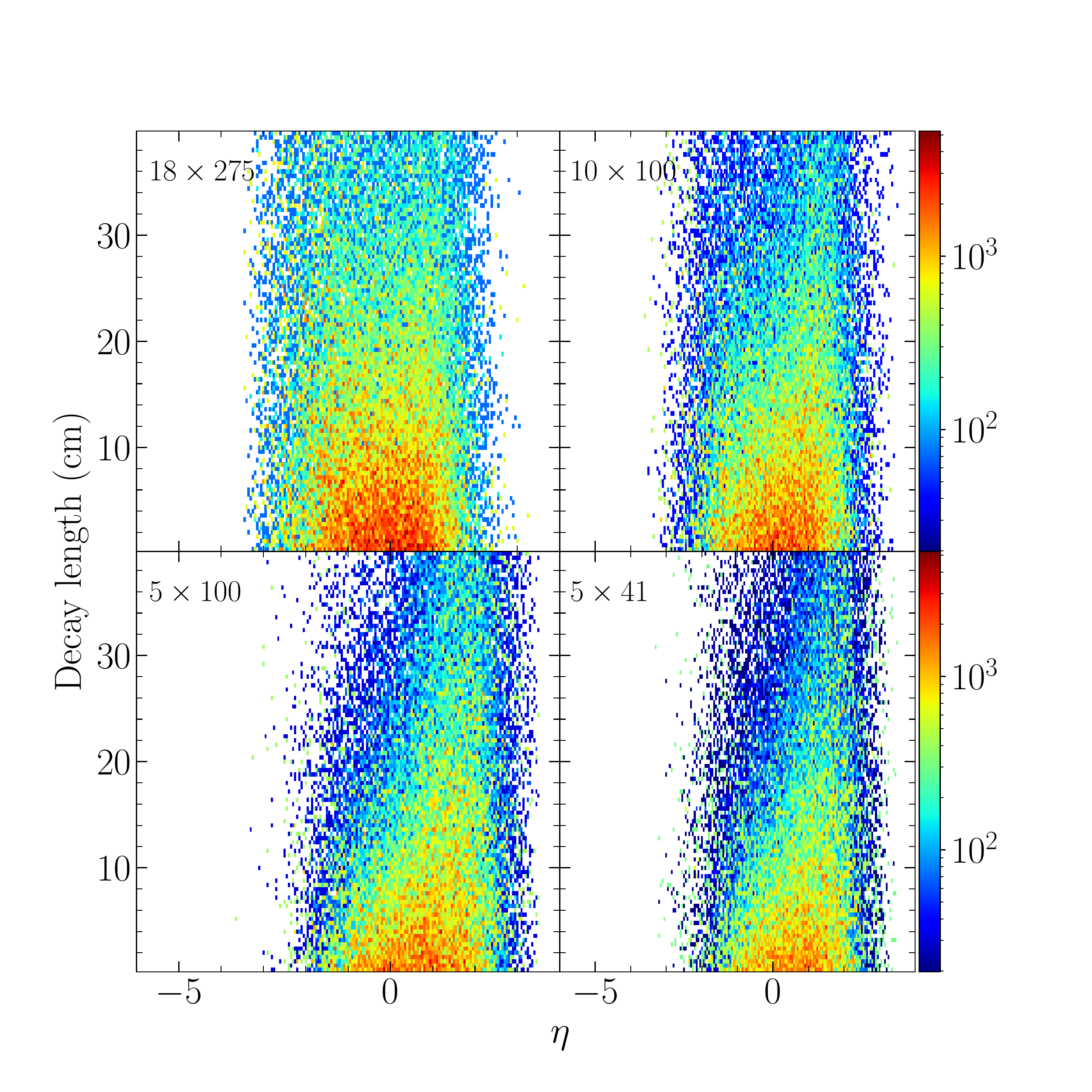}
    \caption{Correlations between $\L$ and $\bar{\L}$ decay length and pseudo-rapidity at different collision energies. }
    \label{fig:lambda_dl_eta}
\end{figure}

Figure~\ref{fig:x_q2} shows the $x_B$ vs. $Q^2$ distribution with the $y$ constraint applied for different collision energies. The $\L$ hyperons are reconstructed from its charged final state decay products, proton (anti-proton), and negative (positive) pion. The 2-D distributions in $p_T$, the transverse momentum in the lab frame, and $\eta$, the pseudo-rapidity space for proton and charged pion, are shown in Fig.~\ref{fig:daughter_pt_eta} (left and middle).  The assumed EIC detector will cover the full azimuth in a finite pseudo-rapidity range $-3.5< \eta< 3.5$. The lowest transverse momentum is set to be $0.1$ GeV.  To ensure the applicability of TMD factorization~\cite{Collins:2011zzd}, the condition, $P_{h \perp}/z < Q/4$ is also applied. The distribution of $P_{h \perp}/z_\Lambda$ versus $Q/4$ is shown in Fig.~\ref{fig:lambda_qtZ} with the dashed line indicating the selection cut. To reduce the contribution from the beam remnant, the Feynman-$x$, $x_F = 2p_L^\L/W$, is required to be positive. As shown in Fig.~\ref{fig:lambda_origins}, the fraction of the $\L$ originating from target fragmentation in the final sample is only a few percent.
This study relies on fast simulations, where the efficiency is not impacted by the displacement between the decay vertex of the hyperon and the primary vertex. As shown in Fig.~\ref{fig:lambda_dl_eta}, the decay vertex can be removed from the primary vertex by several centimeters which might impact the detection efficiency of a compact tracking system proposed for the EIC~\cite{Arrington:2021yeb}. To account for this effect, we apply a quite conservative overall efficiency factor of  $50\%$ for the projected statistical uncertainties.

For $\L$ production at EIC energies, feed-down from heavier particles is not negligible. Figure ~\ref{fig:lambda_origins}, shows the origins of the detected $\L$, according to the event list provided by the PYTHIA generator, for the top energy 18$\times$275 GeV$^2$. After selection cuts, about 1/3$\sim$1/2 of the $\L$  candidates are promptly produced from string fragmentation. 
Most of the remainder, about half of the total, originates from the feed-down of $\Sigma^0$ hyperons, excited $\Sigma^*$ states and $\Xi$ hyperons. Additional contributions, less than 10~\% come from heavy quark decays, {\it e.g.} $\Lambda_c$ and diquarks from the target remnant.
In principle, the fragmentation formalism used in this work, does not apply to hyperons produced in the weak decay of heavier states. This includes most of the feed-down except for feed-down from $\Sigma^*$, which predominantly decays strongly into $\L$. However, to our knowledge, all previous experimental measurements, except for Ref.~\cite{Guan:2018ckx}, did not separate between weak and strong production. Consequently, previous phenomenological work integrated over all $\L$, extracting in some sense effective fragmentation functions. For these reasons, we will also integrate over all $\L$ ancestries and will not assign a systematic uncertainty to the feed-down contributions. 

For an eventual feed-down correction, the contributions from the various decays would have to be identified in data. For $\Sigma^0$, which decays nearly always to $\Lambda+\gamma$, we investigated therefore the feasibility of reconstructing this decay. As shown in Fig.~\ref{fig:daughter_pt_eta} right panel, the final state $\gamma$ are emitted predominantly at or near  the central pseudo-rapidity range with a relative low energy, mostly $<0.5$ GeV. While the detection of these photons is challenging, with the detector performance requirements outlined in the Yellow Report, it should be feasible to reconstruct $\Sigma^0$ hyperons with sufficient mass resolution.  More than half of the feed-down from $\Xi$ hyperons originate from the decay $\Xi^0\rightarrow \Lambda+\pi^0$, which is difficult to reconstruct, due to combinatorial background in the $\pi^0$ reconstruction. The decay $\Xi^\pm\rightarrow \Lambda + \pi^\pm$ can be reconstructed with sufficient efficiency, however, this decay makes up less than 10\%. 
We note that additional systematic uncertainties are expected to come from the uncertainty on the beam polarization for the spin transfer measurement as well as from wrongly reconstructed $\Lambda$ hyperons as well as detector effects. For the $\Lambda$ in jet measurements, the Jet Energy Resolution (JER) will impact the reconstruction of kinematic variables. Following studies in the Yellow Report, the relative uncertainty on the beam polarization can be assumed to be less than 3~\%, which makes this systematic negligible compared with the expected statistical uncertainties. The JER should be about 10\%. Since the extracted quantities are not strongly dependent on kinematics derived from the jet energy, we also assume that this systematic is negligible. Finally, based on experience from previous measurements of $\Lambda$ polarization, we also assume that systematics due to detector effects and wrongly reconstructed $\L$ hyperons are negligible compared to our projected statistical uncertainties. These assumptions have to be revisited for the eventual measurement.

\section{Reweighting Analysis for TMD PFF}\label{sec:Simulation}
Having summarized the details of our experimental simulation in the previous section, we now present our SIDIS re-weighting analysis at the EIC. In Sec.~\ref{subsec:Reweighting-Numerical}, we provide the numerical input for the re-weighting analysis. In Sec.~\ref{subsec:Results}, we provides the results of our projections for the uncertainties for the TMD PFF from this analysis as well as the comparison with the experimental data. 
\subsection{Numerical Input}\label{subsec:Reweighting-Numerical}
For this paper, we will always use LO matching. This is motivated by the fact that the one loop expression for the TMD PFF has not yet been performed. At this perturbative order, the hard function can be replaced by
\begin{align}
    & H^{\rm DIS}(Q) = 1\,.
\end{align}
As a result, the unpolarized structure function and the structure function associated with spontaneous $\L$ polarization in Eqs.~\eqref{eq:strUU} and \eqref{eq:strUT} can be written as
\begin{align}
\label{eq:FUU}
& F_{UU} = \sum_q e_q^2 \int d^2 \bm{k}_{\perp} d^2 \bm{p}_{\perp} \delta^{(2)}(z_\L \bm{k}_{\perp}+\bm{p}_{\perp} -\bm{P}_{h \perp})
\nonumber \\
&\hspace{1cm} \times f_{q/p}(x_B, k_{\perp}^2, Q) D_{\L/q}(z_\L, p_{\perp}^2, Q)\,,
\\
\label{eq:FUT}
&F_{UT}^{\sin(\phi_S - \phi_\L)} = \frac{1}{z_\L M_\L} \sum_q e_q^2 \int d^2 \bm{k}_{\perp} d^2 \bm{p}_{\perp} \\
& \hspace{1cm}\times \delta^{(2)}(z_\L \bm{k}_{\perp}+\bm{p}_{\perp} -\bm{P}_{h \perp})\, \left(\hat{\bf{P}}_{h \perp}\cdot \bm{p}_{\perp}\right)
\nonumber \\
& \hspace{1cm}\times f_{q/p}(x_B, k_{\perp}^2, Q) D_{1T,\L/q}^{\perp}(z_\L, p_{\perp}^2, Q) \nn \,.
\end{align}
In order to parameterize the transverse momentum dependence of the TMDs, we follow the work in Ref.~\cite{Callos:2020qtu} to use a Gaussian parameterization. Using this parameterization, the unpolarized TMDs can be written as
\begin{align}\label{eq.pdfpar}
f_{q/p}(x_{B},k_{\perp}^2, Q)=f_{q/p}(x_{B},Q)\frac{e^{-k_{\perp}^{2}/\langle k_{\perp}^{2} \rangle}}{\pi\langle k_{\perp}^{2}\rangle},
\end{align}
\begin{align}\label{eq:ff_upol_gauss}
D_{\L/q}(z_{\L},p_{\perp}^{2},Q) &=  D_{\L/q}(z_{\L},Q)\; \frac{e^{-p_{\perp}^{2}/\langle p_{\perp}^{2}\rangle}}{\pi \langle p_{\perp}^{2} \rangle}\,,
\end{align}
where $\langle k_{\perp}^{2}\rangle$ and $\langle p_{\perp}^{2}\rangle$ are the Gaussian widths of the TMD PDF and TMD FF, respectively. For this analysis we take $\langle k_{\perp}^{2}\rangle = 0.61$ and $\langle p_{\perp}^{2}\rangle = 0.19$ from Ref.~\cite{Anselmino:2013lza,Callos:2020qtu}.
Furthermore, $f_{q/p}$ and $D_{\L/q}$ are the collinear PDF and FF, respectively. As we will later discuss in Sec.~\ref{sec:SIDIS-Spin-Transfer}, the \texttt{MSTW2008lo68cl}~\cite{Martin:2009iq} parameterization for the collinear PDF is particularly useful for the purposes of this paper. We will then choose to use this parameterization. Additionally, in order to ensure that we are describing the Belle data, we will follow the work in Ref.~\cite{Callos:2020qtu} to use the \texttt{AKK} fragmentation functions~\cite{Albino:2005mv} for the collinear FF.

Using the parameterization in \cite{Callos:2020qtu}, the TMD PFF is given explicitly as
\begin{align}\label{eq:ff_pol_gauss}
\hspace{-2mm}D_{1T,\L/q}^{\perp}(z_{\L},p_{\perp}^2,Q) = D_{1T,\L/q}^{\perp}(z_{\L},Q)
\frac{e^{-p_{\perp}^{2}/\langle M_{D}^{2}\rangle}}{\pi \langle M_{D}^{2} \rangle}\,.
\end{align}
In this expression, $\langle M_{D}^{2} \rangle$ is the Gaussian width of the TMD PFF while the collinear dependence is parameterized as 
\begin{align}
D_{1T,\L/q}^{\perp}(z_{\L},Q) = \mathcal{N}_{q}(z_{\L}) D_{\L/q}(z_{\L},Q)\,.
\end{align}
Here $\mathcal{N}_{q}$ contain the modulation of the TMD PFF, which was parameterized as
\begin{align}
\mathcal{N}_{q}(z_{\L})= N_{q}z_{\L}^{\alpha_{q}}(1-z_{\L})^{\beta_{q}} \frac{(\alpha_q+\beta_q-1)^{\alpha_q+\beta_q-1}}{(\alpha_q-1)^{\alpha_q-1}{\beta_q}^{\beta_q}}\,,
\end{align}
where $N_q$, $\alpha_q$, and $\beta_q$ are fit parameters which were obtained in \cite{Callos:2020qtu}.
At this point it is also convenient to define the first moment of the TMD PFF which is given by
\begin{align}
    D_{1T, \L/q}^{\perp\, (1)}(z_\L,Q) = \frac{\langle M_D^2\rangle}{2 z_\L^2 M_\L^2} D_{1T, \L/q}^{\perp}(z_\L,Q)\,.
\end{align}

Using the parameterizations for the TMDs in Eqs.~\eqref{eq.pdfpar}, \eqref{eq:ff_upol_gauss}, and \eqref{eq:ff_pol_gauss}, the spontaneous $\L$ polarization in Eq.~\eqref{eq:pol-def} can be written as
\begin{align}
\label{eq:Pol-SIDIS-Gauss}
& P_{\L}(x_B,y,z_\L, P_{h \perp}) = \\
& \hspace{1cm} \frac{\sum_q e_q^2 f_{q/p}(x_B,Q) \, \omega_{q}(z_\L,P_{h \perp})
	\, D_{\L/q}(z_{\L},Q)}{\sum_q e_q^2 f_{q/p}(x_B,Q) D_{\L/q}(z_\L,Q)} \,.\nn
\end{align}
In this expression
\begin{align}
\omega_{q}(z_\L,P_{h \perp})& = 
\frac{P_{h \perp}}{z_\L\,M}\frac{\langle M_D^2\rangle \langle P_{h \perp}^2\rangle}{\langle \Delta P_{h \perp}^2\rangle^2}\, \mathcal{N}_q(z_\L) \\
& \times \exp\left[P_{h \perp}^2\left(\frac{1}{\langle P_{h \perp}^2\rangle}-\frac{1}{\langle \Delta P_{h \perp}^2\rangle}\right)\right]\,\nn 
\end{align}
are the weighting functions while 
\begin{align}
\langle P_{h \perp}^2 \rangle = \langle k_\perp^2\rangle z_\L^2+\langle p_\perp^2\rangle\,,
\\
\langle \Delta P_{h \perp}^2 \rangle = \langle k_\perp^2\rangle z_\L^2+\langle M_D^2\rangle\,,
\end{align}
are the Gaussian widths associated with the unpolarized and polarized processes, respectively.

To quantify the contribution of the EIC in constraining the TMD PFFs, we perform two fits in this section. Our baseline fit contains only the experimental data from the Belle collaboration while the re-weighted fit contains both the experimental data from Belle and the pseudo-data generated in the previous section. For this section, we take the integrated luminosity of the EIC pseudo-data to be 40 fb$^{-1}$.

In this analysis, we follow the parameterization in Ref.~\cite{Callos:2020qtu} with fit parameters $N_u$, $N_d$, $N_s$, $N_{sea}$, $\alpha_u$, $\alpha_d$, $\alpha_s$, $\alpha_{sea}$, $\beta_{val}$, $\beta_{sea}$, and $\langle M_D^2\rangle$.  We note that the Belle experimental data was gathered in double inclusive hadron production in $e^++e^-$  annihilation. In this paper, we do not provide the theoretical expression for this polarization and instead refer the reader to our previous work in Ref.~\cite{Callos:2020qtu} which contains this expression as well as the methodology for fitting these data. To perform the fit of the generated pseudo-data, we integrate the numerator and denominator of Eq.~\eqref{eq:Pol-SIDIS-Gauss} in $x$, $y$, and $P_{h\perp}$. Namely to generate the pseudo-data, events are binned into $1>x_B>10^{-1}$, $10^{-1}>x_B>10^{-2}$, $10^{-2}>x_B>10^{-3}$, and $10^{-3}>x_B>10^{-4}$. To generate our theoretical predictions, we integrate over these ranges of $x$ values. pseudo-data are also generated using the constraint that $0.05<y<0.95$. Using the relation $x_B\, y\, S_{\rm ep} = Q^2$, for each data point in our prediction, we integrate over $0.05<y<0.95$ under the condition that $Q>1$ GeV. Finally, to generate the pseudo-data, we have also applied the kinematic constraint that $P_{h\perp}/z_\L<0.25\, Q$, which is associated with the TMD factorization region. For each point, we integrate over this kinematic region in our fitting procedure.

To perform both of the fits, we use the Migrad fit in the \texttt{Minuit} package \cite{1975CoPhC..10..343J} to minimize the $\chi^2$. Furthermore, to generate the theoretical results, we use the replica method \cite{Ball:2008by,Signori:2013mda} with 200 replicas. For each of the replicas, we initialize the fit parameters using a Monte Carlo sampler. 

\subsection{Results}\label{subsec:Results}
\begin{figure}[hbt!]
    \centering
    \includegraphics[width = 0.48\textwidth]{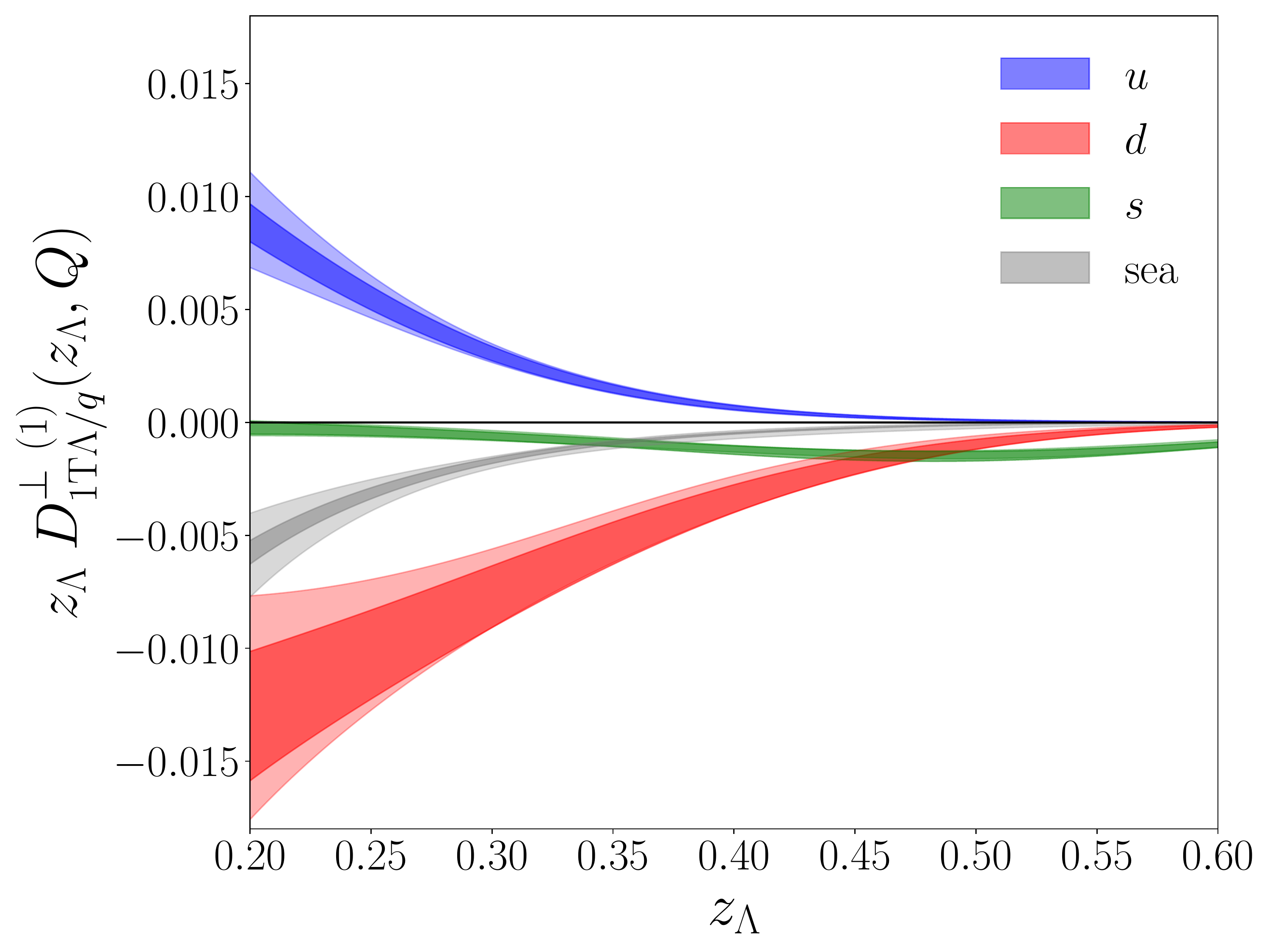}
    \caption{The first moment of the TMD PFF at $Q = 10.58$ GeV. The light bands represent the uncertainty from the fit to Belle data in Ref.~\cite{Callos:2020qtu}, while the dark bands represent the uncertainty obtained from the simultaneous fit of the Belle data and the EIC pseudo-data.}
    \label{fig:TMDPFF}
\end{figure}
\begin{figure*}[hbt!]
    \centering
    \includegraphics[width = 0.8\textwidth]{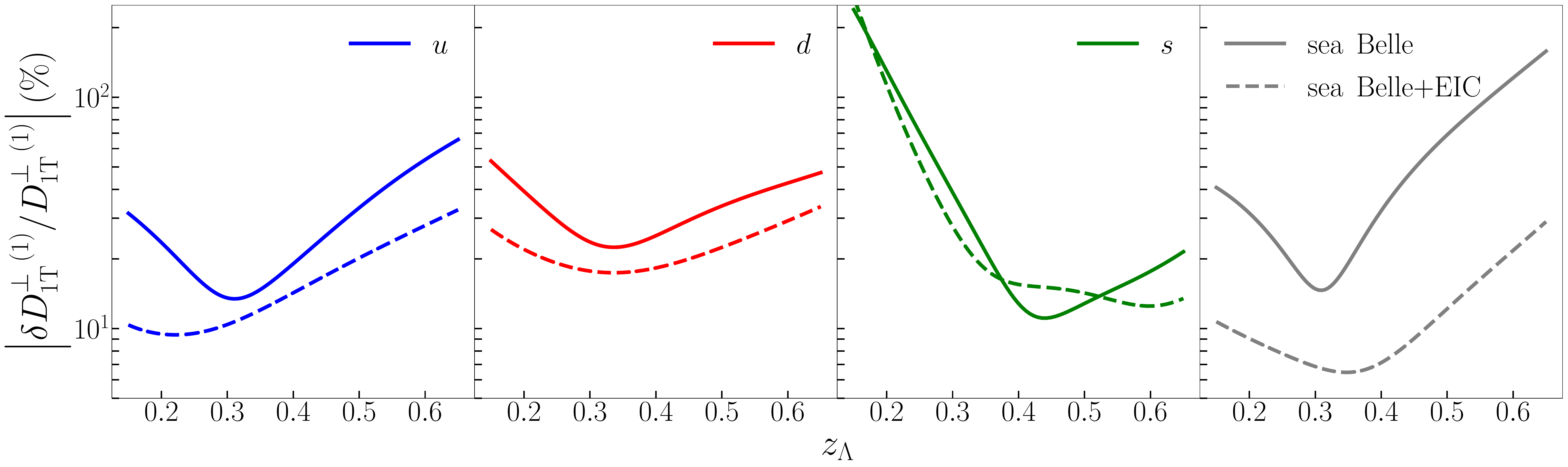}
    \caption{The ratio of the uncertainty of the TMD PFF for each flavor at $Q = 10.58$ GeV. The solid lines represent the results from the fit to Belle data while the dashed line represents the result from the fit to the Belle data and the EIC pseudo-data. }
    \label{fig:Uncertainty}
\end{figure*}
\begin{figure*}[hbt!]
    \centering
    \includegraphics[width = \textwidth]{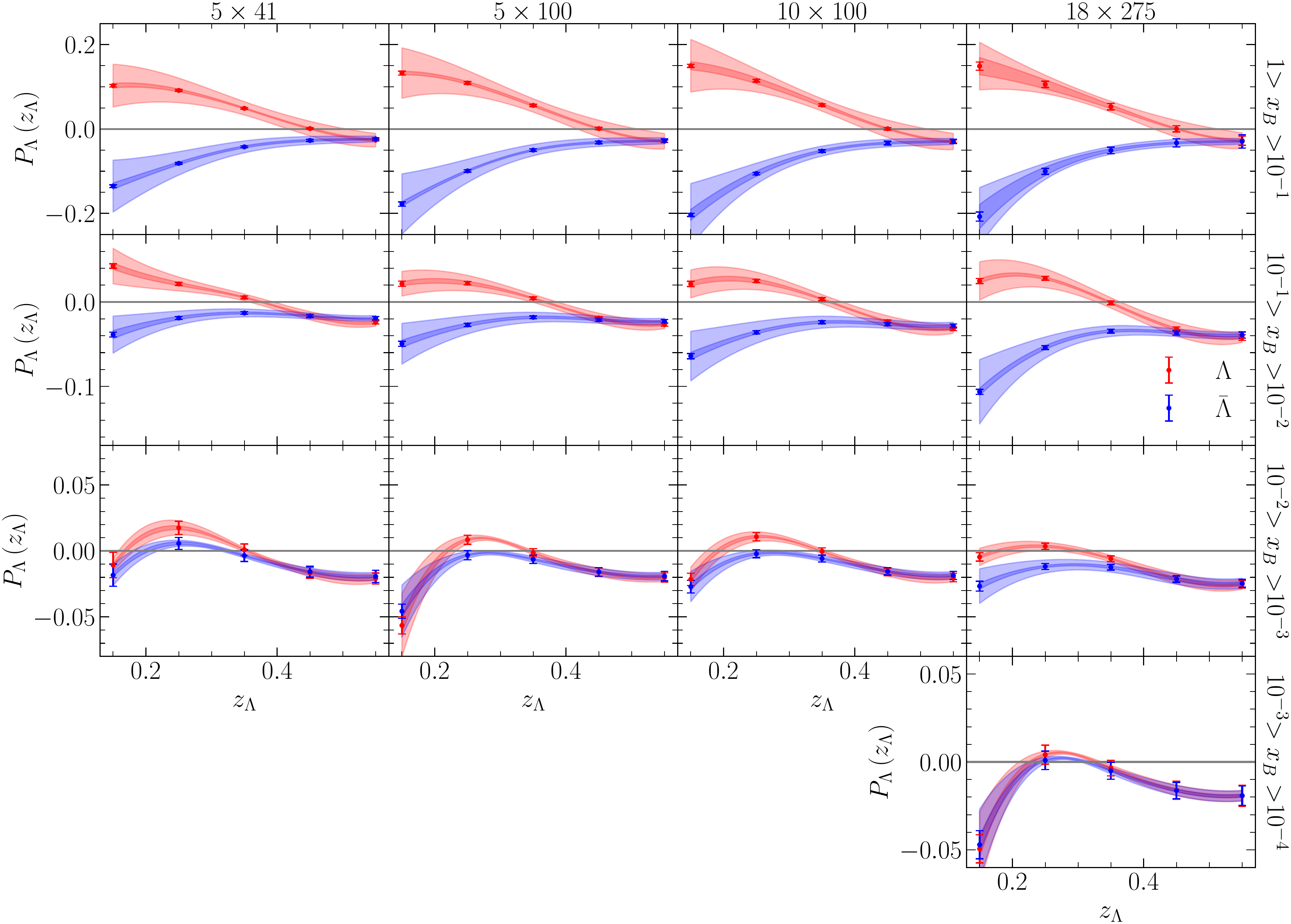}
    \caption{Theoretical predictions are compared against the pseudo-data generated in the reweighting method at 40 fb$^{-1}$. The rows are grouped by the range of $x_B$ while the columns are group by the energy configuration of the collision where the first number represents the energy of the lepton beam in GeV and the second number represents the energy of the hadron beam in GeV. The light band represents the theoretical uncertainty from the baseline fit to the Belle data while the dark band represents the theoretical uncertainty from the fit including the pseudo-data. }
    \label{fig:data}
\end{figure*}
In Fig.~\ref{fig:TMDPFF}, we plot the first moment of the TMD PFF which was obtained from the baseline fit as a light band. In the darker band, we plot the result from the simultaneous fit to the Belle data as well as the EIC pseudo-data. The theoretical uncertainty for the first moment, $\delta D_{1T}^{\perp\, (1)}$, is obtained from the set of replicas by calculating the standard deviation at each point. Furthermore, we define the average value of the extracted first moment as $D_{1T}^{\perp\, (1)}$. As we can see, from this plot, the uncertainty is significantly reduced in the simultaneous analysis. In order to further quantify the size of this reduction of the theoretical uncertainties, we also plot the ratio $\delta D_{1T}^{\perp\, (1)}/D_{1T}^{\perp\, (1)}$ in Fig.~\ref{fig:Uncertainty}. As we see in this figure, the pseudo-data generated from the EIC kinematics leads to a significant reduction of the theoretical uncertainty for the $u$ and $\rm sea$ quarks. The large reduction in the theoretical uncertainties in the sea quark TMD PFF is occurring because the parameterization in Ref.~\cite{Callos:2020qtu} assumes charge symmetry. Thus $D_{1T,\L/q}^{\perp\,(1)}$ = $D_{1T,\bar{\L}/\bar{q}}^{\perp\,(1)}$ and $D_{1T,\L/\bar{q}}^{\perp\,(1)}$ = $D_{1T,\bar{\L}/q}^{\perp\,(1)}$. Since the analysis that we perform here uses a proton beam, the fit with the pseudo-data allows us to strongly constrain the $D_{1T,\bar{\L}/u}^{\perp\,(1)}$ and $D_{1T,\bar{\L}/d}^{\perp\,(1)}$ functions. From the charge symmetry assumption, this leads to a large reduction in the uncertainties for the $D_{1T,\L/\bar{u}}^{\perp\,(1)}$ and $D_{1T,\L/\bar{d}}^{\perp\,(1)}$ functions. As a result, the uncertainties for the sea quark distributions are dramatically reduced. However, because the strange distribution in the PDF is small, we find that the theoretical uncertainties $D_{1T,\L/s}^{\perp\,(1)}$ are unchanged with the introduction of pseudo-data. Furthermore, we find that the reduction in the theoretical uncertainties for the $d$ quark distribution is smaller than those for the $u$ and sea quarks. However, we note that in principle the theoretical uncertainty for this flavor can be further reduced by considering experimental data from a $^{3}$He nuclear beam. Namely by using a $^{3}$He beam and tagging the two protons in the forward region, a neutron can be isolated in the hadron beam. This procedure would allow future extractions to spontaneous $\L$ production for electron-neutron scattering which is extremely useful for studying the $d$-quark TMD PFF.

In Fig.~\ref{fig:data}, we plot the pseudo-data that was obtained in the previous section for $\L$ in red and $\bar{\L}$ in blue. Once again, the light band represents the theoretical uncertainty obtained from the baseline fit. The dark band represents the theoretical uncertainty from the simultaneous fit. To calculate the theoretical uncertainties, we compute the standard deviation of replicas at each point. We bin each row (column) of this plot according to its $x_B$ range (energy configuration). The top row contains the pseudo-data generated from events with $1>x_B>10^{-1}$. As we can see in this region, the size of the projected polarization is relatively large and positive for $\L$ production. Furthermore, the polarization is large and negative for the $\bar{\L}$ production. The large and positive polarization for $\L$ production is occurring because the contribution from the $u$ quark is dominant for this process and is also positive. The large and negative polarization for $\bar{\L}$ production is occurring because the large contribution from the $u$ quark TMD PDF is being weighted with the sea contribution for the TMD PFF. Since the sea contribution is negative, the resulting asymmetry is large and negative. As the binned value of $x_B$ decreases as we move down the rows in Fig.~\ref{fig:data}, we can see that the polarization for $\L$ tends to decrease in magnitude. This behavior occurs because there are large cancellations between the $u$ and sea quarks in this kinematic region. 

\section{Spin transfer in SIDIS}\label{sec:SIDIS-Spin-Transfer}
In this section, we study the potential role of the future EIC in constraining the transversity TMD FF. In order to characterize the theoretical uncertainty which can be obtained by current experimental data, we first use the current experimental data from COMPASS to constrain the transversity TMD FF for $\L$ baryon production. Using the theoretical uncertainty obtained from this extraction, we generate a prediction at EIC kinematics and compare the projected theoretical uncertainties against the projected statistical uncertainties.

In this section, we first begin by providing the numerical recipe used for the extraction of the transversity TMD FF. To perform this analysis, we work at LO accuracy for the matching and Next-to-Leading-Logarithmic (NLL) accuracy for the logarithmic resummation. Beyond the Gaussian approximation, it is convenient to work in $b$-space. The expressions for the unpolarized structure function in Eq.~\eqref{eq.strFUU} at LO becomes
\begin{align}
    F_{UU} = \sum_q e_q^2 \int & \frac{db\, b}{2\pi} \,J_0\left(\frac{bP_{h \perp}}{z_\L}\right) \\
    \times & D_{\L/q}(z_\L,b, Q) f_{q/p}(x_B,b, Q)\,.  \nn
\end{align}
At the perturbative accuracy that we use in this section, the TMDs can be matched onto the collinear distributions using the relations
\begin{align}
    \label{eq:TMDPDF-matching}
    & f_{q/p}(x_B,b,Q) = f_{q/p}(x_B,\mu_{b_*}) \\
    & \hspace{2cm} \times \exp\left(-S_{\rm pert}(\mu_{b_*},Q) -S_{\rm NP}^f(b,Q)\right)\,,\nn
    \\
    \label{eq:TMDFF-matching}
    & D_{\L/p}(z_\L,b,Q) = \frac{1}{z_\L^2}D_{\L/q}(z_\L,\mu_{b_*}) \\
    & \hspace{2cm} \times \exp\left(-S_{\rm pert}(\mu_{b_*},Q) -S_{\rm NP}^D(z_\L,b,Q)\right) \nn\,.
\end{align}
In these expressions, $f_{q/p}$ and $D_{\L/q}$ are the collinear PDF and FF, respectively. In order to arrive at this expression, we have used the $b_*$ prescription from \cite{Collins:1984kg} with $b_{\rm max} = 1.5$ GeV$^{-1}$. Furthermore, $S_{\rm pert}$ represents the perturbative Sudakov factor which is responsible for evolving the distribution from $\mu_{b_*}$, the natural scale, to the hard scale $Q$. The expression for this function is given for instance in Ref.~\cite{Echevarria:2020hpy} by 
\begin{align}
S_{\rm pert}(\mu_{b_*},Q) =& -\tilde K(b_*,\mu_{b_*})\ln\left(\frac{Q}{\mu_{b_*}}\right) \\& -\int_{\mu_{b_{*}}}^{Q} \frac{d\mu}{\mu}\left[\gamma_F\left(\alpha_s, \frac{Q^2}{\mu^2}\right) \right] \nn \,.
\end{align}
At NLL order one has $K(b_*,\mu_{b_*})=0$ and 
\begin{align}
    & \gamma_F\left(\alpha_s, \frac{Q^2}{\mu^2}\right) = \gamma_{\rm cusp}\left(\alpha_s\right)\, \ln\frac{Q^2}{\mu^2}+\gamma_V\left(\alpha_s\right)\,,
\end{align}
where
\begin{align}
    & \gamma_{\rm cusp} (\alpha_s) = \frac{\alpha_s}{4\pi}\left(4C_F\right) 
    \label{eq:cusp}
    \\
    & \hspace{5mm}+\left(\frac{\alpha_s}{4\pi}\right)^2\left\{ 4C_F \left[C_A\left(\frac{67}{9}-\frac{\pi^2}{3}\right)-\frac{20}{9} T_R\, n_f\right]\right\}\,,\nn
\\
    &\gamma_V(\alpha_s) = \frac{\alpha_s}{4\pi}\left(-6 C_F\right)\,.
\end{align}
are the cusp and non-cusp anomalous dimensions. 

The $S_{\rm NP}$ functions in Eqs.~\eqref{eq:TMDPDF-matching} and \eqref{eq:TMDFF-matching} are the non-perturbative Sudakov functions. For this analysis, we follow the parameterization in~\cite{Echevarria:2020hpy,Su:2014wpa}
\begin{align}
    \label{eq:TMDPDF-NP-parameterization}
    S_{\rm NP}^f(b, Q) = g_q\,b^2 + \frac{g_2}{2}\, \ln\left(\frac{b}{b_*}\right)\ln\left(\frac{Q}{Q_0}\right)\,,
    \\
    \label{eq:TMDFF-NP-parameterization}
    S_{\rm NP}^D(z_\L,b, Q) = \frac{g_h}{z^2}b^2+\frac{g_2}{2}\, \ln\left(\frac{b}{b_*}\right)\ln\left(\frac{Q}{Q_0}\right)\,,
\end{align}
where $g_q = 0.106$ GeV$^2$, $g_h = 0.042$ GeV$^2$, and $g_2 = 0.84$ and $Q_0^2 = 2.4$ GeV$^2$. The unpolarized structure function in Eq.~\eqref{eq:strUU} can be written as
\begin{align}
& F_{UU} = \sum_q e_q^2 \frac{1}{z_\L^2} \int \frac{db\, b}{2\pi} J_0\left(\frac{bP_{h \perp}}{z_\L}\right)\\
& \times D_{\L/q}(z_\L, \mu_{b_*})\, f_{q/p}(x_B, \mu_{b_*}) \nn \\
& \times \exp\left[-2 S_{\rm pert}(\mu_{b_*},Q)-S_{\rm NP}^D(z_\L,b,Q)-S_{\rm NP}^f(b,Q)\right] \nn \,.
\end{align}

\begin{figure*}[hbt!]
    \centering
    \includegraphics[width = 0.8\textwidth]{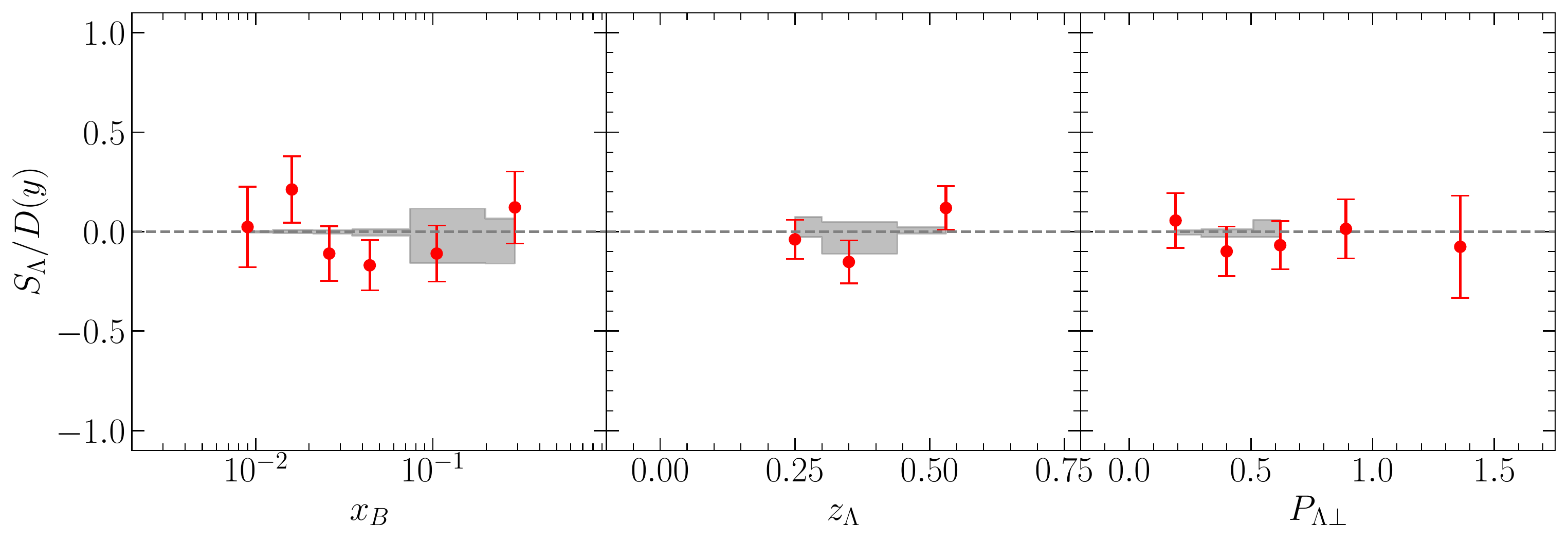}
    \caption{The comparison of our fitted transverse spin transfer and the experimental data at COMPASS~\cite{Alexeev:2021aws}. The grey band represents our theoretical uncertainty which is obtained using the replica method, while the red error bars are the experimental data from COMPASS.}
    \label{fig:COMPASS}
\end{figure*}

\begin{figure}[hbt!]
    \centering
    \includegraphics[width = 0.48\textwidth]{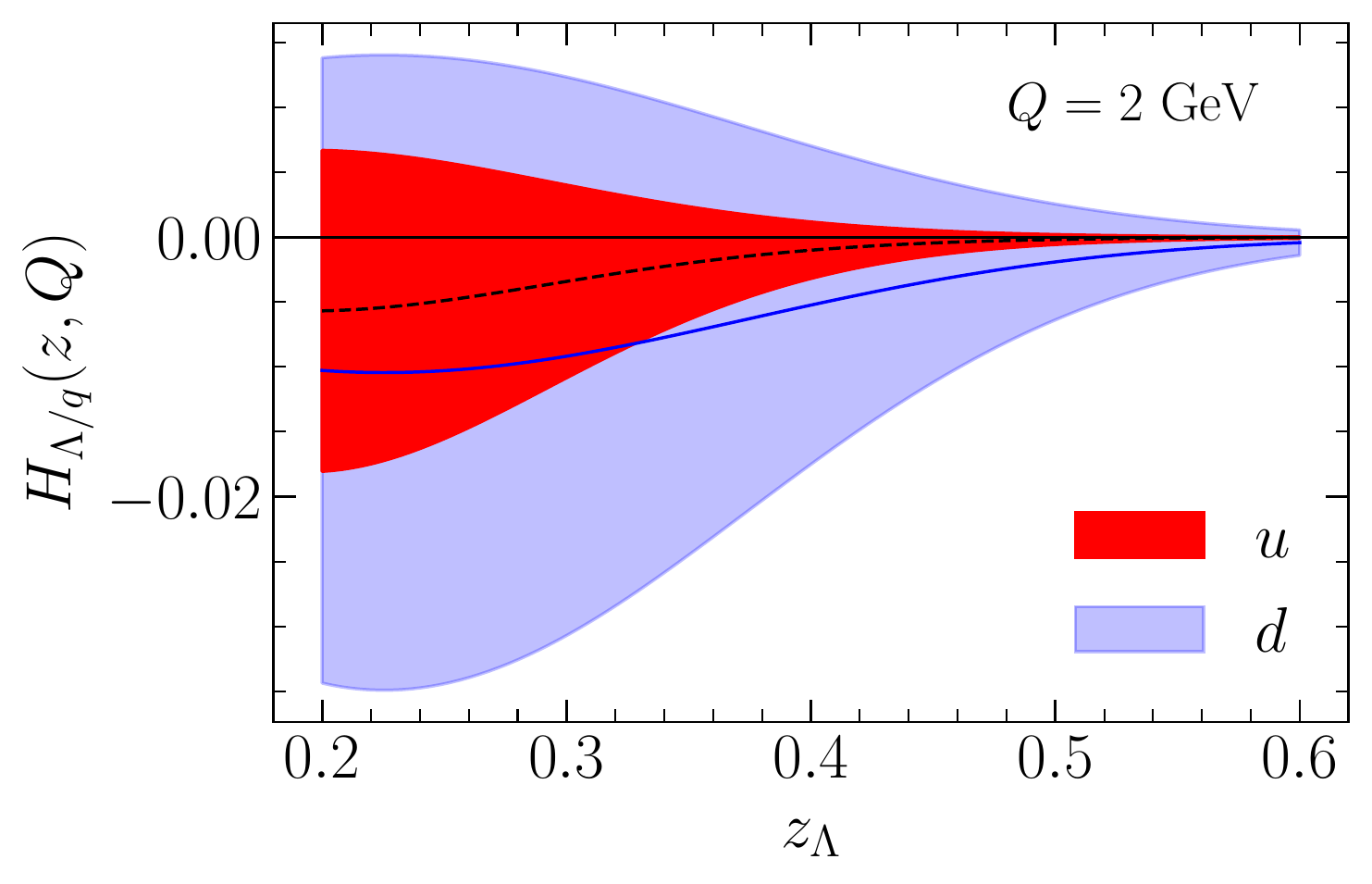}
    \caption{The collinear transversity FF extracted from the COMPASS measurement. The red and blue bands represent the theoretical uncertainties from our extraction which was obtained using the replica method for the $u$ and $d$ quarks, respectively. The dashed and solid lines represent the average of the replicas for the $u$ and $d$ quarks.}
    \label{fig:Trans-FF}
\end{figure}
Following a similar analysis for the transverse spin transfer process, one can write the structure function associated with this process as
\begin{align}
\label{eq:FTT}
& F_{TT}^{\cos(\varphi_S - \phi_S)} = \sum_q e_q^2 \frac{1}{z_\L^2} \int \frac{db\, b}{2\pi} J_0\left(\frac{bP_{h \perp}}{z_\L}\right)\\
& \times H_{\L/q}(z_\L, \mu_{b_*})\, h_{q/p}(x_B, \mu_{b_*}) \nn \\
& \times \exp\left[-2 S_{\rm pert}(\mu_{b_*},Q)-S_{\rm NP}^H(z_\L,b,Q)-S_{\rm NP}^h(b,Q)\right] \nn
\end{align}
where $h_{q/p}$ and $H_{\L/q}$ are the collinear transversity PDF and FF while $S_{\rm NP}^H$ and $S_{\rm NP}^h$ are the non-perturbative Sudakov factors for these distributions. 

Several extraction in the literature for the transversity TMD PDF. In this work, we follow the work in Ref.~\cite{Kang:2015msa} to parameterize the transversity PDF as
\begin{align}\label{eq:trans-param}
    h_{q/p}(x, \mu_0) & = N_q^h x^{\alpha_q^h}(1-x)^{\beta_q^h} \frac{(\alpha_q^h+\beta_q^h)^{\alpha_q^h+\beta_q^h}}{{\alpha_q^h}^{\alpha_q^h}{\beta_q^h}^{\beta_q^h}} \nn \\
    & \times \frac{1}{2}\left[ f_{q/p}(x,\mu_0)+g_{q/p}(x,\mu_0)  \right] \,.
\end{align}
In this expression, $N_q^h$, $\alpha_q^h$, $\beta_q^h$ are fit parameters which were obtained in this reference for the $u$ and $d$ quarks while the contributions of the sea quarks were set to zero. Furthermore, we have defined the initial scale of the parameterization to be $\mu_0$. 

The transversity PDF in Eq.~\eqref{eq:trans-param} can in general have a non-trivial $x$ dependence. This non-trivial dependence enters because, while the collinear PDF $f$ and the helicity PDF $g$ have simply polynomial dependencies on $x$ at their initial scales, when these initial scales differ from $\mu_0$, evolution effects in $f$ and $g$ will complicate this parameterization. The simplest parameterization the one could take would be to choose the scale $\mu_0$ such that it corresponds to the intrinsic scales of both $f$ and $g$. This constraint limits the number of parameterizations that we can use for these distributions. While there are a large number of parameterization for $f$ available on the market, there are relatively few parameterization for $g$. In this analysis, we take the DSSV parameterization from Ref.~\cite{deFlorian:2009vb}. The initial scale of this parameterization is $1$ GeV. Of the available PDFs, we find that the MSTW parameterization shares the same initial scale as the DSSV parameterization. We have therefore chosen to use this set for the entire paper. As a result of this choice, the $x$ dependence of our parameterization of the transversity PDF can be shown to be given by a polynomial at $\mu_0$.

In order to evolve $h_{q/p}$ from the initial scale to $\mu_{b_*}$ in Eq.~\eqref{eq:FTT}, we must solve the DGLAP evolution equation for this distribution. However, as was stated in Ref.~\cite{Stratmann:2001pt}, there is no gluon transversity at leading power. As a result, the DGLAP evolution equation of the quark transversity does not mix with the gluon distribution. Therefore, the evolution equation does not contain splitting function which mix quarks and gluons and the DGLAP evolution equation is simply given by
\begin{align}\label{eq:evo-trans}
    \frac{\partial }{\partial\, \ln \mu^2} h_{q/p}(x, \mu) = \frac{\alpha_s}{2\pi} \int_x^1 \frac{d\hat{x}}{\hat{x}} P_{q\rightarrow q}^{h}(\hat{x}) \, h_{q/p}\left(\frac{x}{\hat{x}}, \mu\right)\,,
\end{align}
where the splitting kernel for the transversity PDF is given by
\begin{align}
    P_{q\rightarrow q}^{h}(x) = C_F \left[\frac{2 \hat{x}}{(1-\hat{x})_+}+\frac{3}{2}\delta(1-\hat{x})\right]\,.
\end{align}
The evolution equation in Eq.~\eqref{eq:evo-trans} can be drastically simplified by taking the Mellin transform of this expression. Because our parameterization for the transversity PDF resulted in polynomial dependence on $x$ at $\mu_0 = 1$ GeV, the Mellin transform for the transversity PDF can be performed analytically at $\mu_0 = 1$ GeV. As a result, evolving our parameterization for the transversity PDF from $\mu_0$ to $\mu_{b_*}$ can be accomplished by performing a single numerical integral which is associated with an inverse Mellin transformation.

To parameterize the transverse momentum dependence on the transversity TMD PDF, we follow the parameterization in Ref.~\cite{Kang:2015msa}. Explicitly, we parameterize the non-perturbative Sudakov for the transversity TMD PDF as
\begin{align}
    S_{\rm NP}^h(b,Q) = S_{\rm NP}^f(b,Q)\,,
\end{align}
which sets the non-perturbative Sudakov to be the same for unpolarized TMD PDF and the transversity TMD PDF.

Having parameterized the transversity TMD PDF, we now turn our attention to the transversity TMD FF. As the COMPASS measurement is consistent with zero~\cite{Alexeev:2021aws}, these experimental data can provide relatively little input on the size and shape of $H_{\L/q}$. Therefore, we choose the relatively simple parameterization
\begin{align}
    H_{\L/q}(z, Q) & = N_q^H\,D_{\L/q}(z,Q)\,,
\end{align}
where $N_q^H$ represents parameters to be fit which control the overall size of the transversity FF. 

Because the parameterization for the transversity TMD PDF in Eq.~\eqref{eq:trans-param} has only non-zero contributions from the $u$ and $d$ quarks, only the $N_u^H$ and $N_d^H$ parameters can be constrained in this analysis. As such, we take all other quark contributions to be zero. Due to these assumptions, our model will predict zero transverse spin transfer for $\bar{\L}$ production. Therefore, we do not consider the $\bar{\L}$ production data for this process. 

To parameterize the non-perturbative Sudakov term for the transversity TMD FF, we follow the procedure that was done in in Ref.~\cite{Kang:2015msa} to set the non-perturbative Sudakov term to be the same for the unpolarized TMD PDF and the transversity TMD PDF. For the transversity TMD FF, we explicitly take
\begin{align}
S_{\rm NP}^H(z_\L,b,Q) = S_{\rm NP}^D(z_\L,b,Q)\,.
\end{align}
\begin{figure*}[hbt!]
    \centering
    \includegraphics[width = \textwidth]{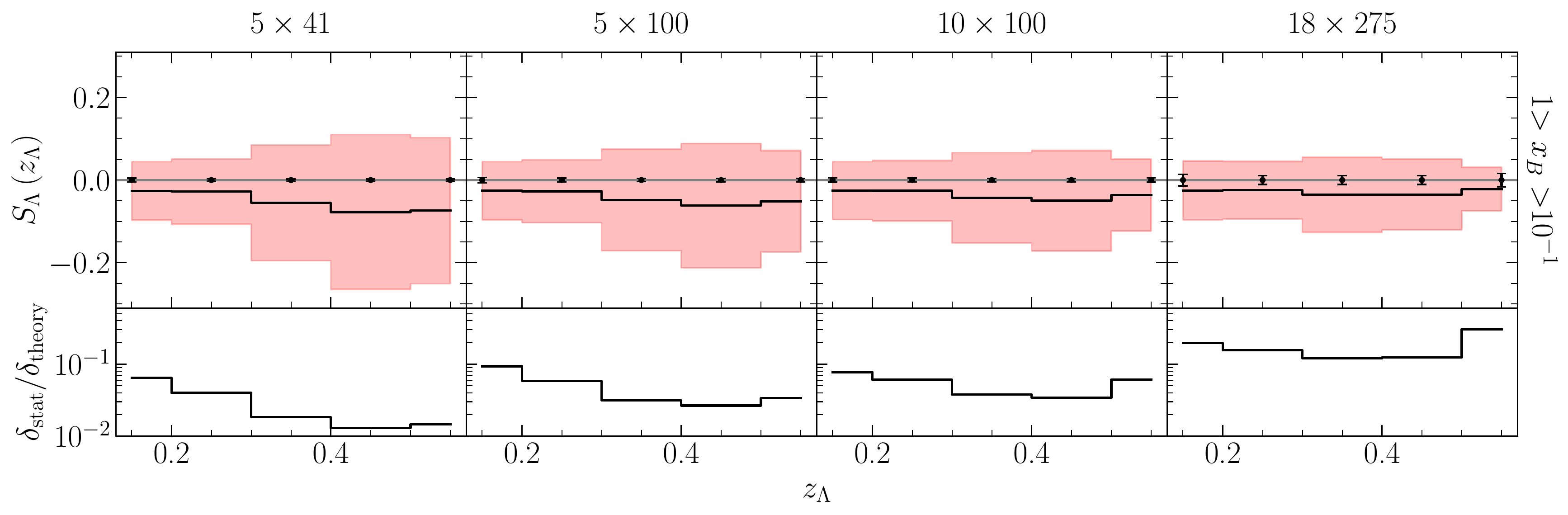}
    \caption{Prediction for the transverse spin transfer for $\L$ production at the future EIC. The label above each sub plot provides the electron beam energy $\times$ the proton beam energy. The label to the right indicates the kinematic region for $x_B$. In each subplot, the red bars represents the theoretical uncertainty from our extraction at one standard deviation. The black line represents the average over the replicas. The black error bars represent the projected statistical uncertainties at 40 fb$^{-1}$.}
    \label{fig:dataTT}
\end{figure*}

In order to fit the $N_u^H$ and $N_d^H$ parameters, we use iMinuit \cite{1975CoPhC..10..343J,iminuit}. Furthermore, in order to generate the uncertainty band from the extraction, we use the replica method \cite{Ball:2008by,Signori:2013mda} with 200 replicas. Using this simple model, we arrive at a $\chi^2/d.o.f = 1.108$ for 12 points. The fitted values for the parameters are given by $N_u^H = -0.028\pm 0.061$ and $N_d^H = -0.089\pm 0.210$. 

In Fig.~\ref{fig:COMPASS}, we plot our theoretical comparison against the COMPASS experimental data. The grey band represents our theoretical uncertainty which is obtained by calculating one standard deviation away from the mean of the replicas at each point. As we see in this plot, in each kinematic region, our extraction for the transverse spin transfer is consistent with zero. In Fig.~\ref{fig:Trans-FF}, we plot the extracted collinear transversity FF as a function of $z_\L$ for $Q = 2$ GeV for the $u$ quark in red and the $d$ quark in blue. The dashed and dotted lines represent the average over the replicas for the $u$ and the $d$ quark transversity FFs, respectively. Due to the experimental measurement being consistent with zero, we see that the transversity FF is also consistent with zero within our theoretical error bars.

Having performed this first extraction of the transversity TMD FF, we now compare the theoretical uncertainties on transverse spin transfer from our extraction against the projected statistical uncertainties at the future EIC. In Fig.~\ref{fig:dataTT}, we plot our theoretical prediction at the future EIC for the transverse spin transfer for $\L$ production in SIDIS in the large $x$ region where the valence quarks should dominate. The red band represents the theoretical uncertainty from our extraction while the black line line represents the average of the replicas. The black error bars represent our projected statistical uncertainties at the future EIC. In order to obtain the statistical uncertainties for these kinematic ranges, we have divided the statistical uncertainties from the spontaneous polarization section by a factor of $70\%$ in order to account for the uncertainties associated with the proton beam polarization. We note that the theoretical uncertainty which we display in this analysis stems only from the parameters $N_u^H$ and $N_d^H$ parameters. The full theoretical uncertainties should also contain contributions from the uncertainties from the transversity TMD PDF as well as the unpolarized TMDs and even the collinear distributions. As a result, these theoretical uncertainties underestimate the total theoretical uncertainty. However, as we see in this figure, the theoretical uncertainties are more than an order of magnitude larger than the projected statistical uncertainties at the EIC. This indicates that the EIC could potentially be used to perform the first measurement of the transverse spin transfer which is not consistent with zero and that such data would be extremely important in constraining the transversity TMD FF. 

\section{Projections for $\L$ in Jet}\label{sec:In-Jet}
In this section, we first present our parameterization for spontaneous $\L$ polarization as well as the transverse spin transfer for $\L$ baryon production within the produced jet. We then present the results of our projections at the future EIC. 

The definition of spontaneous $\L$ polarization is given in Eq.~\eqref{eq:pol-in-jet-def}. This expression relies on the unpolarized structure functions $W_{UU}$ in Eq.~\eqref{eq:unp-jet-L}, and the structure function for spontaneous polarization $W_{UT}^{\sin(\phi_S - \phi_\L)}$ in Eq.~\eqref{eq:pol-jet-L}. At this point, we first provide the parameterization for the unpolarized structure function. 

In this section, we once again work at LO+NLL perturbative order. At LO matching, the hard functions in Eq.~\eqref{eq:unp-jet-L} and Eq.~\eqref{eq:trans-jet-L} are
\begin{align}
    H(Q,\mu) = 1\,, \qquad H_\perp(Q,\mu) = 1\,.
\end{align}
For this process, there are two separate soft function which contribute to $\bm{q}_\perp$. The first contribution is the well-known global soft function, which we will denote $S_{\rm global}$. This function is associated with wide angle soft gluon emissions. The second contributions is known as the collinear soft function, which we denote $S_{\rm sc}$, is associated with soft gluon exiting the jet, see for instance Ref.~\cite{Buffing:2018ggv} for more details. The global soft and collinear-soft functions for this process are given up to NLO+NLL accuracy in Ref.~\cite{Kang:2021ffh} along with the anomalous dimensions. At LO+NLL accuracy, these functions are given by
\begin{align}
    S_{\rm global}(b, \mu) = \exp\left[ \int_{\mu_{b_*}}^\mu \frac{d\mu'}{\mu'}\gamma_{\rm global}(b, \mu')  \right]\,,
\end{align}
\begin{align}
    S_{\rm sc}(b, y_J, R, \mu) = \exp\left[ \int_{\mu_{b_*}}^\mu \frac{d\mu'}{\mu'}\gamma_{\rm sc}(b, y_J, R,\mu')  \right]\,,
\end{align}
where the anomalous dimensions for these functions are given at NLL accuracy by
\begin{align}
    \gamma_{\rm global}(b, \mu) =&\, 2\frac{\alpha_s C_F}{\pi} y_J+\gamma_{\rm cusp} C_F \ln\frac{\mu^2}{\mu_b^2}\,,
\\
    \gamma_{\rm sc}(b, y_J, R, \mu) =&\, -\gamma_{\rm cusp} C_F\, \ln\frac{\mu^2}{\mu_b^2 R^2}\,.
\end{align}
with $\mu_b = 2 e^{-\gamma_E}/b$. In the $b$-space, the soft functions combine as a product so that the total soft function entering into this process, i.e. $U(b,y_{J},R,\mu)$ in Eqs.~\eqref{eq:unp-jet-L}, \eqref{eq:pol-jet-L} and \eqref{eq:trans-jet-L}, is given by
\begin{align}
    U(b,y_{J},R,\mu) = S_{\rm global}(b, \mu) S_{\rm sc}(b,y_J, R,\mu)\,.
\end{align}
In our numerical analysis, we will always take the jet radius $R = 1$ so that the anomalous dimension of $U$ is simply given by the sum of the global and collinear-soft anomalous dimensions.

In order to obtain the final expression for the unpolarized structure functions in Eq.~\eqref{eq:unp-jet-L}, we now use the collinear matching expression in Eq.~\eqref{eq:TMDFF-matching} to write the unpolarized TMD fragmenting jet function as
\begin{align}
    & {\cal G}_{\L/q}(z_{J \L},j_\perp,\mu_J,\mu)=\exp\left[\int_{\mu_J}^{\mu} \frac{d\mu'}{\mu'} \gamma_J(\mu') \right] \\
    & \hspace{1cm}\times \int\frac{db\,b}{2\pi}J_0\left( \frac{b j_\perp}{z_{J \L}}\right) D_{\L/q}(z_{J \L}, \mu_{b_*})\nn \\
    & \hspace{1cm}\times \exp\left[- S_{\rm pert}(\mu_{b_*},\mu_J)-S_{\rm NP}^D(z_{J \L},b,\mu_J)\right] \nn \,.
\end{align}
At NLL, the anomalous dimension for the TMD fragmenting jet function is given by
\begin{align}
    \gamma_J(\mu) =  -\gamma_{\rm cusp}(\alpha_s) \ln\left(\frac{\mu_J^2}{\mu^2}\right)-\gamma_V(\alpha_s) \,.
\end{align}
In addition, we also include contributions from the non-global logarithms, see Refs.~\cite{Dasgupta:2001sh,Liu:2018trl,Buffing:2018ggv,Banfi:2003jj,Chien:2019gyf} for details. Finally, in order to obtain the structure function for unpolarized $\L$ production, we also apply the matching relation for the TMD PDF in Eq.~\eqref{eq:TMDPDF-matching} onto the expression in Eq.~\eqref{eq:unp-jet-L}. After performing the matching, the unpolarized structure function is given by
\begin{align}
&W_{UU} = \,\sum_q e_q^2\, {\cal G}_{\L/q}(z_{J \L},j_\perp,\mu_J,\mu)
    \\ &\times \,
\int\frac{db\, b}{(2\pi)}\, J_0(b\,q_\perp)\, f_{q/p}(x_B,\mu_{b_*})\,U(b,y_{J},R,\mu) \nn\\
    & \times \exp\left(-S_{\rm pert}(\mu_{b_*},\mu)-S_{\rm NP}^f(b,\mu)\right) \,. \nn
\end{align}

\begin{figure}[bt!]
    \centering
    \includegraphics[width = 0.48\textwidth]{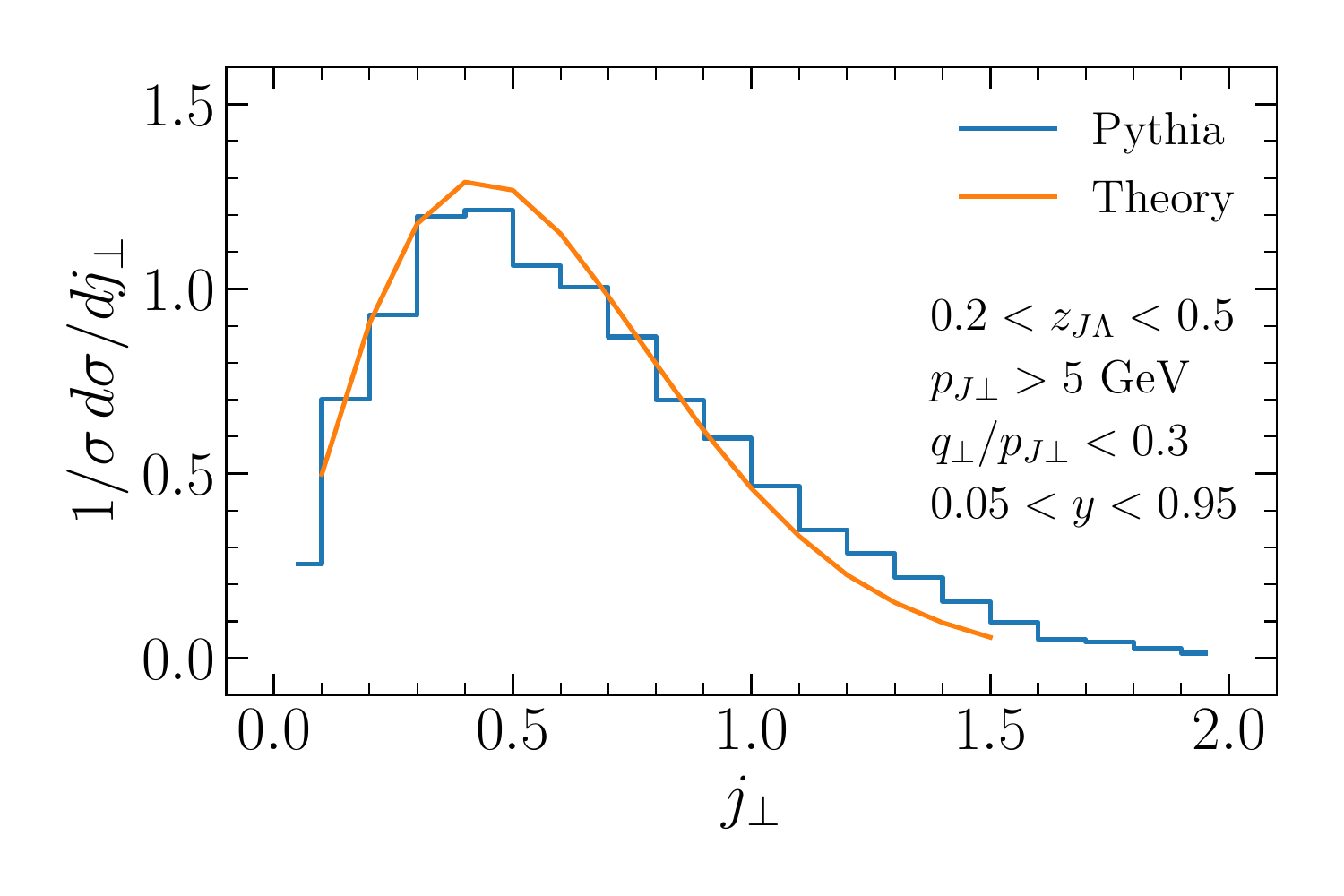}
    \caption{The $j_\perp$ distribution for unpolarized $\L$ baryons in a jet for back-to-back lepton-jet production in $ep$ collisions. The orange line represents our theoretical prediction while the blue histogram represents the Pythia simulation. The integrated phase space is also displayed on the right side of this figure.}
    \label{fig:upol-JT}
\end{figure}
To obtain numerical results for this section, we use the same parameterization for the unpolarized TMDs as in Sec.~\ref{sec:SIDIS-Spin-Transfer}. 

In order to verify the validity of our formalism so far, we have included a comparison with Pythia in Fig.~\ref{fig:upol-JT}. In this figure, we plot our $j_\perp$ distribution for unpolarized $\L$ production inside the jet. In our Monte Carlo analysis where we generated the pseudo-data for this process, we have examined events which satisfy the constraints $0.05<y<0.95$, $p_{J\perp}>5$ GeV, $q_\perp/p_{J\perp}<0.3$, and $0.2<z_{J \L}<0.5$. Therefore, in order to generate our theoretical prediction for this data, we integrate the structure functions entering into the polarization over these kinematic regions. To perform the integration in $y$, we simply use the relation in Eq.~\eqref{eq:yey} to relate the lepton rapidity to the inelasticity. To perform the integration in $p_{J\perp}$, we note that up to power corrections of $q_\perp/p_{j\perp}$ that $\ell'_\perp = p_{j\perp}$ so that we can simply perform the integration in the jet transverse momentum. We have also taken $R = 1$ for the jet radius. In this figure, the Pythia histogram as well as our theoretical curve have been normalized by integrating over $j_\perp< 1.5$ GeV. As we can see in this figure, the shape of the $j_\perp$ distribution matches the result of the Pythia simulation extremely well. 

Using the collinear matching relation for the TMD PDF, the structure function for spontaneous $\L$ polarization can be written as 
\begin{align}
&W_{UT}^{\sin(\phi_S-\phi_\L)} = \,\sum_q e_q^2\, {\cal G}_{1T,\L/q}^{\perp}(z_{J \L},j_\perp,\mu_J,\mu)
    \\ &\times \,
\int\frac{db\, b}{(2\pi)}\, J_0(b\,q_\perp)\, f_{q/p}(x_B,\mu_{b_*})\,U(b,y_{J},R,\mu) \nn\\
    & \times \exp\left(-S_{\rm pert}(\mu_{b_*},\mu) -S_{\rm NP}^f(b,\mu)\right) \,. \nn
\end{align}
In order to simplify the TMD polarizing fragmenting jet function, we introduce the collinear matching relation for the TMD PFF
\begin{align}
    \label{eq:matching-PFF}
 D_{1T,\L/q}^{\perp(1)}\left(z,b,Q\right)&= \frac{\langle M_D^2\rangle}{2z^2 M_\L^2}
D_{1T,\L/q}^{\perp}  (z,\mu_{b_*})\\
    &\times e^{ - S_{\rm pert}(\mu_{b_*},Q)-S_{\rm NP}^\perp(b,z,Q)} \nn \,.
\end{align}
Using this collinear matching relation, the TMD fragmenting jet function can be written as
\begin{align}
    & {\cal G}_{1T,\L/q}^{\perp}(z_{J \L},j_\perp,\mu_J,\mu)=-\frac{\langle M_D^2\rangle}{2z_{J \L}^4 M_\L}\sin\left(\phi_s-\phi_\L\right)
    \\
    & \hspace{1cm} \times \exp\left[\int_{\mu_J}^{\mu} \frac{d\mu'}{\mu'} \gamma_J(\mu') \right] \int\frac{db\, b}{2\pi}J_1\left(\frac{b j_\perp}{z_{J \L}}\right) \nn
    \\
    & \hspace{1cm} \times \, D_{1T, \L/q}^\perp(z_{J \L}, \mu_{b_*}) e^{-S_{\rm pert}(\mu_{b_*},\mu_J)-S_{\rm NP}^\perp (z_{J \L},b,\mu_J)} \nn \,.
\end{align}
\begin{figure*}[hbt!]
    \centering
    \includegraphics[width = \textwidth]{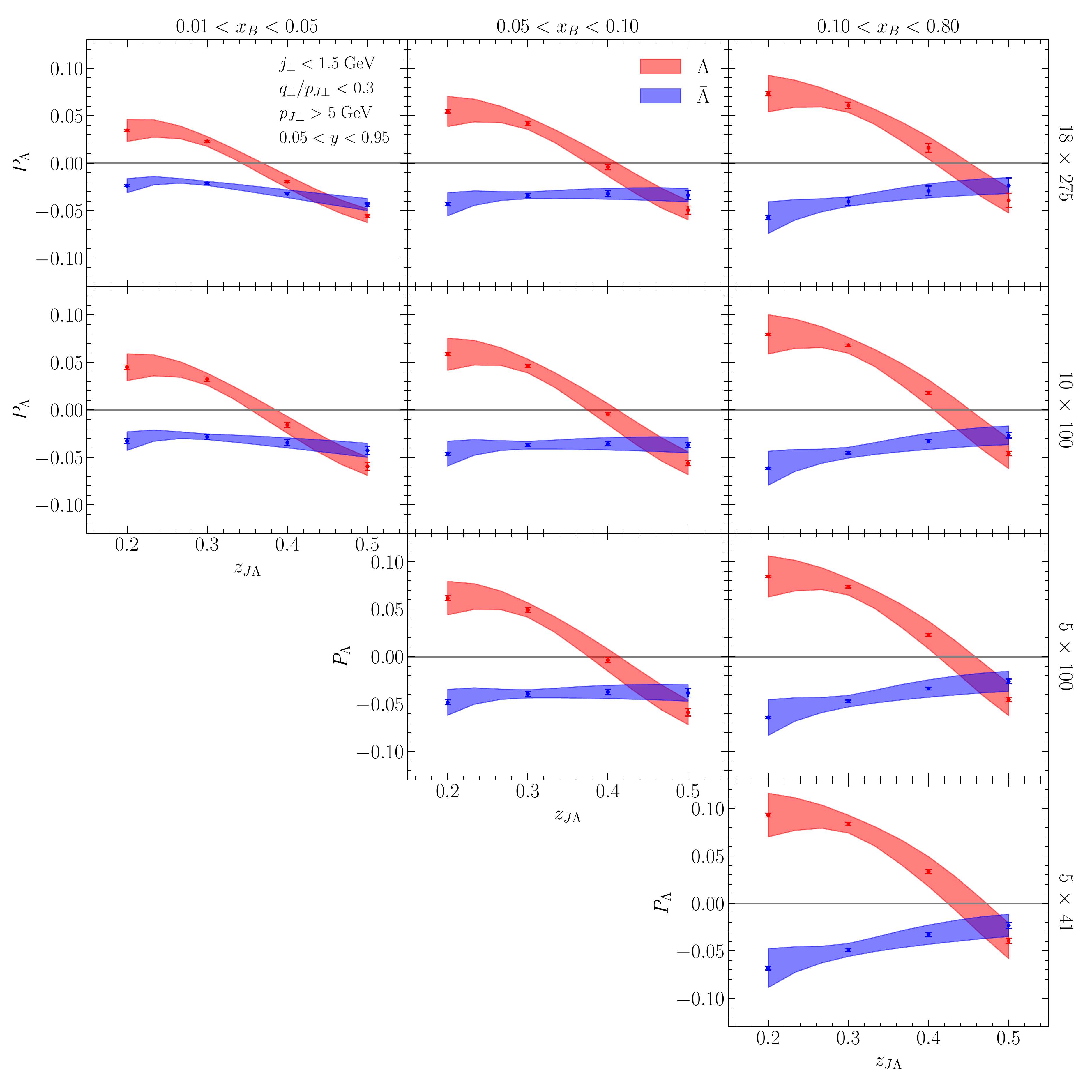}
    \caption{Our projections for spontaneous $\L$ polarization. The red band and blue bands represents our theoretical uncertainty using the parameters obtained from the baseline fit in Sec.~\ref{subsec:Results} for $\L$ and $\bar{\L}$ production, respectively. The error bars represent our projected statistical uncertainties at an integrated luminosity of $100$ fb$^{-1}$. To obtain these results, we integrate over the kinematic regions listed in the top right of the left plot. Furthermore, we also impose the conditions that $x_B$ is within each of the listed regions.}
    \label{fig:Lambda-In-Jet}
\end{figure*}

In Fig.~\ref{fig:Lambda-In-Jet}, we plot our theoretical prediction for spontaneous $\L$ polarization for back-to-back electron-jet production. The red and blue curves represent the theoretical uncertainty for $\L$ and $\bar{\L}$ production in which we obtain from the baseline fit in Sec.~\ref{subsec:Results}. The red and blue error bars represent the statistical uncertainties for $\L$ and $\bar{\L}$ production at an integrated luminosity of 100 fb$^{-1}$. To generate each curve, we integrate over the kinematic region $j_\perp< 1.5$ GeV, $q_\perp/p_{J\perp}<0.3$, $p_{J\perp}>5$ GeV, and $0.05<y<0.95$ following the same procedure as in the unpolarized case. From left to right, we impose the kinematic constraint that $0.01<x_B<0.05$, $0.05<x_B<0.10$, and $0.10<x_B<0.80$. In each of these plots, we see that the polarization for $\L$ is positive at small $z_{J \L}$, while the polarization becomes negative at large $z_{J \L}$. Furthermore, we also find that the polarization for $\L$ is more positive at small $z_{J \L}$ and large $x_B$. These qualitative behaviors can be seen by studying Fig.~\ref{fig:TMDPFF}. At small $z_{J \L}$, the contribution from the $u$ quark will dominate the polarization due to the electro-magnetic coupling of the $u$ quark as well as the size of the $u$ quark TMD PDF. As a result, the polarization is large and positive at small $z_{J \L}$. At large $z_{J \L}$ the contributions from the other quark flavors overcome the $u$ quark and the polarization becomes negative. Since the contribution from the $u$ quark is largest in the large $x_B$ region, the polarization is more positive at large $x_B$. For $\bar{\L}$ production, the $u$ and $d$ are sea contributions to the TMD PFF. As a result, the contributions from the $u$ and $d$ quarks give large negative contributions to the polarization. We see in these plots that the size of the statistical uncertainties is smaller than the theoretical uncertainties in the region of small $x_B$. This is an indication that experimental data gathered in that particular region can be useful in further constraining the TMD PFF. However, the displayed theoretical uncertainties stem only from the the uncertainties from the fit parameters for the TMD PFF. Other theoretical uncertainties stemming from the unpolarized TMD PDF as well as the collinear distributions will also contribute to this prediction.

\begin{figure*}[hbt!]
    \centering
    \includegraphics[width = \textwidth]{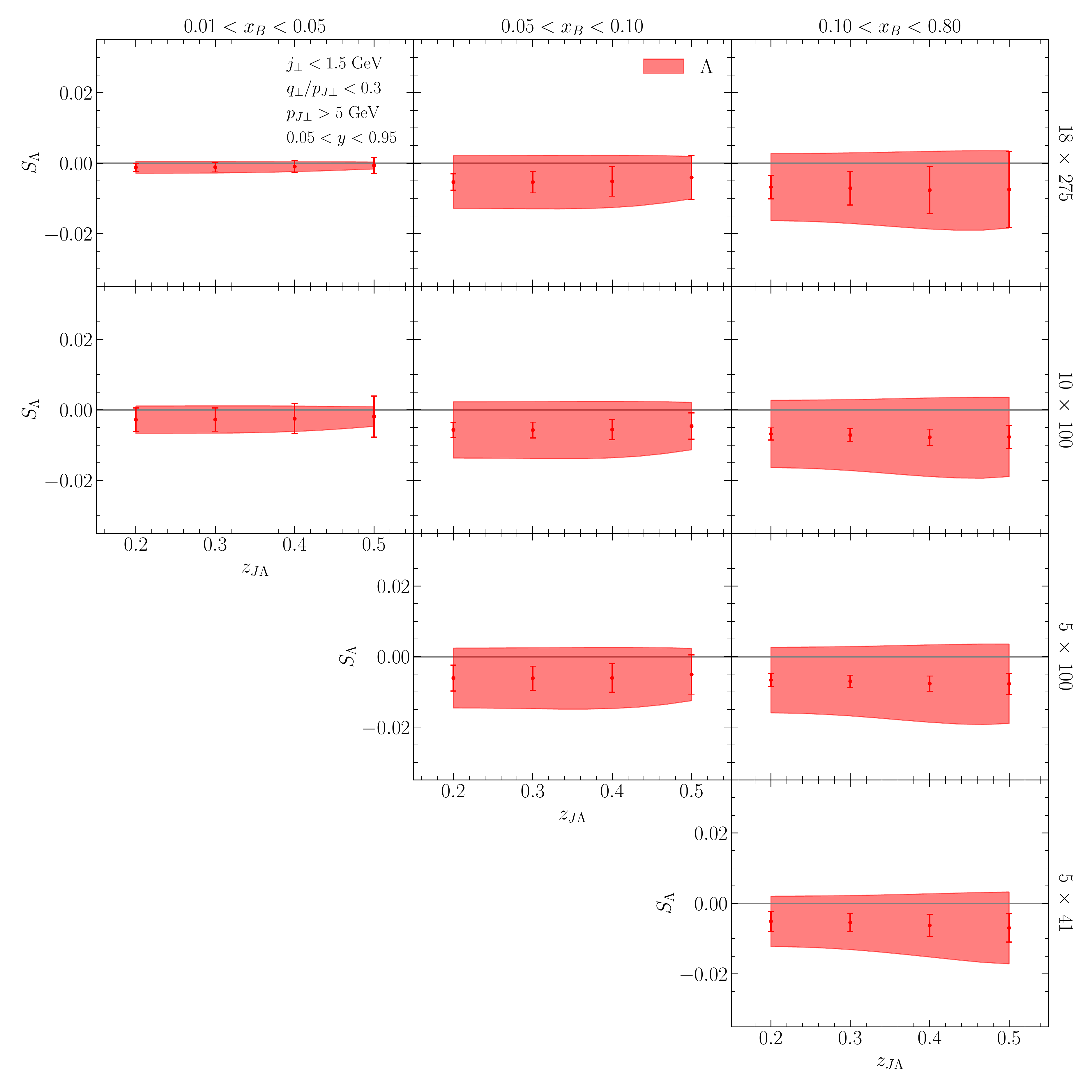}
    \caption{Our projection for the transverse spin transfer for $\L$ production for back-to-back lepton-jet production at the future EIC. The red bar represents our theoretical uncertainties which we obtain from our extraction of the $N_u^H$ and $N_d^H$ parameters. The error bars represent the projected statistical uncertainties at 100 fb$^{-1}$. We have obtained these statistical uncertainties by dividing the uncertainties from Fig.~\ref{fig:Lambda-In-Jet} by $70\%$ to account for beam polarization uncertainty.}
    \label{fig:Trans-In-Jet}
\end{figure*}

After performing the collinear matching for the transversity TMD PDF, the structure function associated with the transverse spin transfer is given by
\begin{align}
&W_{TT}^{\cos\left(\varphi_{S}-\phi_{S}\right)} = \sum_q e_q^2\, {\cal G}_{\L/q}^T(z_{J \L},j_\perp,\mu_J,\mu)
    \\ &\times \,
\int\frac{db\, b}{2\pi}\, J_0(b\,q_\perp)\, h_{q/p}(x_B,\mu_{b_*}) U(b,y_{J},R,\mu) \nn\\
    & \times \exp\left(-S_{\rm pert}(\mu_{b_*},\mu)-S_{\rm NP}^h(b,\mu)\right) \,. \nonumber
\end{align}
In this expression, the TMD fragmenting jet function is given by
\begin{align}
    {\cal G}_{\L/q}^T&(z_{J \L},j_\perp,\mu_J,\mu)=\exp\left[\int_{\mu_J}^{\mu} \frac{d\mu'}{\mu'} \gamma_J(\mu') \right] \\
    & \frac{1}{z_{J \L}^2}\int\frac{db\, b}{2\pi}J_0\left(\frac{b j_\perp}{z_{J \L}}\right)\, H_{\L/q}(z_{J \L},\mu_{b_*}) \nn\\
    & \times \exp\left(-S_{\rm pert}(\mu_{b_*},\mu_J)-S_{\rm NP}^H(z_{J \L},b,\mu_J)\right) \,. \nonumber \,
\end{align}
To generate our theoretical prediction for back-to-back lepton-jet production, we use the extracted transversity FF from Sec.~\ref{sec:SIDIS-Spin-Transfer} while we once again use the parameterization from Ref.~\cite{Kang:2015msa} for the transversity TMD PDF. 

In Fig.~\ref{fig:Trans-In-Jet}, we plot our projected transverse spin transfer in the region of large $x_B$ where the contribution from the valence quarks should dominate. The red bar represents the theoretical uncertainty for our fit to the $N_u^H$ and $N_d^H$ parameters while the error bar is the projected statistical uncertainty. To generate the statistical uncertainties for this measurement, we use the statistical uncertainties used in Fig.~\ref{fig:Lambda-In-Jet}, while dividing by a factor of $70\%$ to account for the uncertainty in the polarization of the proton beam. We once again emphasize that the advertised theoretical uncertainty stems only from the parameters that enter into our fit while we expect additional large uncertainties originating from the transversity TMD PDF, the unpolarized TMDs, as well as the unpolarized collinear distributions also contribute to this measurement. 

\section{Conclusion}\label{sec:Conclusion}
In this paper we have studied $\L$ production at the future EIC for spontaneous transverse $\L$ polarization as well as transverse spin transfer in the TMD formalism. Furthermore, we have studied each of these spin configurations in SIDIS as well as back-to-back lepton-jet production. For each of these processes, we have discussed the impact of the future EIC in constraining the TMD PFF as well as the transversity TMD FF.

In order to characterize the size of the contribution that the future EIC data will have on constraining the TMD PFF, we have performed an EIC impact study. As a baseline we have performed a fit to the experimental data at Belle. While in order to test the impact of the EIC data, we have performed a Pythia analysis to generate projections for the statistical uncertainties at the future EIC. Using these statistical uncertainties, we have performed a simultaneous fit to the Belle data as well as the pseudo-data. By performing this fit, we have demonstrated a significant reduction in the theoretical uncertainties for the $u$ and sea TMD PFF. We have also discussed how potential measurements with a $^3\rm He$ beam can be used to significantly reduce the uncertainty for the $d$ TMD PFF.

In order to study the impact of future EIC data on the transversity TMD FF, we have performed an extraction of this function from the recent COMPASS measurement~\cite{Alexeev:2021aws} in the SIDIS process. We find that the current statistical precision from the COMPASS measurement is not high enough for an extraction of the transversity TMD FF. By providing projections for the statistical uncertainties for $\L$ polarization in the SIDIS process at the future EIC, we demonstrate that the statistical uncertainties for this process at the future EIC will be roughly an order of magnitude smaller than the current theoretical uncertainties for this process. Thus, the EIC data presents the possibility of being the first significant measurement of the transversity TMD FF.

In addition, we have provided projections for $\L$ in jet production in back-to-back lepton-jet production. We have generated projected statistical uncertainties at the future EIC for spontaneous $\L$ production at an integrated luminosity of $100$ fb$^{-1}$. We find that in the region of low $x_B$ that the statistical precision for this process can be used to further constrain the TMD PFF. Finally, we have also provided projections for the transverse spin transfer for $\L$ in jets in the scattering of an electron and a transversely polarized proton at the future EIC, and we emphasize its importance in constraining the transversity TMD FF.

\section*{Acknowledgements}
The authors thank Fanyi Zhao and Kyle Lee for providing Fig.~1. Z.K. is supported by the National Science Foundation under Grant No.~PHY-1945471. J.T. is supported by NSF Graduate  Research Fellowship Program under Grant No.~DGE-1650604 and UCLA Dissertation Year Fellowship. A.V. is supported by the U.S. Department of Energy, Office of Science, Office of Nuclear Physics under Award No.~DE-SC0019230 and No.~DE-AC05-06OR23177.
Q.X. is supported by the National Natural Science Foundation of China under Grant No.~12075140. J.Z. is supported by the Qilu Youth Scholar Funding of Shandong University. This work is supported within the framework of the TMD Topical Collaboration.

 \bibliography{refs}
\end{document}